\documentclass[useAMS,usenatbib]{mn2e}
\usepackage{rotating}
\usepackage{graphics}
\usepackage{graphicx, subfigure}
\usepackage{tabularx}
\usepackage{amssymb}
\usepackage{setspace}
\usepackage{txfonts}
\usepackage{natbib}
\usepackage{longtable}
\usepackage{epstopdf} 
\usepackage{epsfig}
\usepackage{float}
\usepackage{subfig}
\usepackage{tikz, graphicx}
\usepackage{array}
\usepackage{arydshln}

\title[VLA observations of UCDs]{Modelling the environment around five ultracool dwarfs via the radio domain}
\author[ ]
{Metodieva, Y. T.$^{1}$\thanks{e-mail: ytm@arm.ac.uk}, Kuznetsov, A. A.$^{2}$,  Antonova, A. E.$^{3}$, Doyle, J. G.$^{1}$, \and
Ramsay, G.$^{1}$, Wu, K.$^{4}$\\
$^{1}$Armagh Observatory, College Hill, Armagh, BT61 9DG, N. Ireland, UK\\
$^{2}$Institute of Solar-Terrestrial Physics, Irkutsk 664033, Russia\\
$^{3}$Department of Astronomy, Faculty of Physics, St Kliment Ohridski University of Sofia, 5 James Bourchier Boulevard, 1164 Sofia, Bulgaria\\
$^{4}$Mullard Space Science Laboratory, University College London, Holmbury St. Mary, Dorking, Surrey RH5 6NT, UK}

\begin{document}

\date{01 Jan 1900}

\pagerange{\pageref{firstpage}--\pageref{lastpage}} \pubyear{9999}

\maketitle

\label{firstpage}

\begin{abstract}

We present the results of a series of short radio observations of six ultracool dwarfs made using the upgraded VLA in S (2--4GHz) and C (4--7GHz) bands. LSR J1835+3259 exhibits a 100 percent right-hand circularly polarised burst which shows intense narrowband features with a fast \textit{negative} frequency drift of about $-30$ MHz $\textrm{s}^{-1}$. They are superimposed on a fainter broadband emission feature with a total duration of about 20 minutes, bandwidth of about 1 GHz, centred at about 3.5 GHz, and a slow \textit{positive} frequency drift of about 1 MHz $\textrm{s}^{-1}$. This makes it the first such event detected below 4 GHz and the first one exhibiting both positive and negative frequency drifts. Polarised radio emission is also seen in 2MASS J00361617+1821104 and NLTT 33370, while LP 349-25 and TVLM 513-46546 have unpolarised emission and BRI B0021-0214 was not detected. We can reproduce the main characteristics of the burst from LSR J1835+3259 using a model describing the magnetic field of the dwarf as a tilted dipole. We also analyse the origins of the quiescent radio emission and estimate the required parameters of the magnetic field and energetic electrons. Although our results are non-unique, we find a set of models which agree well with the observations.\

\end{abstract}


%-------------------------------------

\section{Introduction}\

In the Sun and other Solar-type stars magnetic fields appear to be ubiquitous, originating within the star's convective envelope through a dynamo process \citep{charbonneau97}. In contrast, stars less massive than 0.35M$_{\odot}$ (around spectral type M4) \citep{chabrier97, stassun11} were expected to be fully convective and therefore, not have a significant large scale magnetic field (see e.g. \citet{reiners07}). However, observations of late-type dwarf stars at optical and other wavelengths show flares which are taken as evidence for the presence of a magnetic field. For instance, the archetypal flare star UV Cet \citep{moffett74} shows many flares despite its M6V spectral type. Another common signature of magnetic activity in late type stars is the presence of H$\alpha$ emission. Optical spectra obtained using the Sloan Digital Sky Survey (SDSS) and the Baryon Oscillation Sky Survey (BOSS) data indicate that by spectral type $\sim$M9V-L0V, approximately 100 per cent of stars show H$\alpha$ emission \citep{west08, schmidt15} with activity continuing in L and T spectral classes \citep{gizis00, fleming03, schmidt07}.\

Despite the evidence for activity in late M type stars in the optical band, it was still a huge surprise to find quiescent and flare-like {\it radio} emission from an M9V dwarf \citep{berger01}.  Further observations show that a small, but significant fraction of low-mass stars with spectral type later than M7 (known as {\sl ultracool dwarfs}, or UCDs, \citet{kirkpatrick99}) show periodic bursts of radio emission which can be up to 100 per cent circularly polarised \citep{hallinan06,berger09,doyle10,tony13,kao15, williams2-15}. These polarised pulses are thought to be produced by the electron cyclotron maser instability (ECMI) \citep{hallinan08}.  Many UCDs which do not show pulsed radio emission are still detected as radio sources. The quiescent radio/X-ray flux ratio from the pulsed UCDs violates the standard G\"{u}del-Benz relation \citep{guedel93} by several orders of magnitude \citep{berger01, williams14}. This quiescent radio emission is generally attributed to gyrosynchrotron radiation \citep{berger02}. However, an alternative explanation is that depolarisation and steady particle acceleration could cause ECMI emission to have low variability \citep{hallinan08}.\

There are strong grounds to believe that the process for generating large scale magnetic fields in stars with spectral type later than $\sim$M4V is quite different to the dynamo process found in the higher mass M dwarfs. Alternative means for generating magnetic fields in low mass stars has received much attention \citep{chabrier06, browning08, morin08, morin10, kitchatinov14, shulyak15}. For instance, \citet{christensen09} extended a scaling law derived from geodynamo models to rapidly rotating stars that have strong density stratification and find sufficient energy flux available for generating the magnetic field. Magnetic fields can, therefore, be generated in small and large planets {\it and} low mass stars. However, the mechanism for generating fields in late M dwarfs is far from settled.\

Determining the magnetic field geometry of a star has traditionally been done using Zeeman Doppler Imaging and lines which are sensitive to magnetic fields in the optical and infrared wavelengths \citep{semel89, donati89}. However, for later spectral types, stars can be rapidly rotating which causes the spectral lines to broaden and blend with each other. They also become faint in the optical bands and spectroscopy in the infrared wavelengths becomes necessary. \

The fact that UCDs can show strong polarised radio pulses, opens up an alternative means to constrain the magnetic field strength and geometry by modelling the pulse profiles at different frequencies. To create the framework to model polarised bursts from UCDs, \citet{yu12} simulated the micro-physics of the immediate environment around late type dwarfs. Other work, such as \citet{kuznetsov12} and \citet{hallinan15}, have created models which take physically realistic environmental conditions coupled with different geometries for the magnetic field to simulate individual polarised radio bursts. For instance, \citet{kuznetsov12} showed that a model of emission from an active sector in the UCD TVLM 513-46546 (hereafter TVLM 513) is able to reproduce qualitatively the main features of its radio light curves; the magnetic dipole seems to be highly tilted by about 60$^{\circ}$ with respect to the rotation axis.\

Existing observations of UCDs have been made primarily with the pre-upgraded Very Large Array (VLA) and the Arecibo 305m dish.  Most of the radio observations of UCDs published prior to 2013 were taken at a single frequency, although there were some exceptions, e.g. data taken using the Arecibo telescope \citep{osten06, osten08, kuznetsov12}. With upgrades to the VLA and Australian Telescope Compact Array (ATCA), giving sensitivity gains of an order of magnitude higher and the opportunity to obtain dynamic spectra over a 2 GHz frequency range, it is possible to observe the dynamical characteristics of polarised radio bursts from UCDs in a way which was previously not possible.  In this paper we present a series of short VLA observations of six UCDs and outline model simulations which have been used to model their quiescent emission as well as a polarised burst from one UCD - LSR J1835+3259.\

\begin{table*}
\begin{minipage}{180mm}
\caption{The sample of dwarfs selected for the VLA observations. Columns show spectral types, masses, periods, distances and known radio flux densities from the literature (from non-flaring emission), and the frequency of observations: 1 -- \citet{berger05}, ~2 -- \citet{berger06},~3 -- \citet{chabrier00},~4 -- \citet{dieterich14},~5 -- \citet{doyle10},~6 -- \citet{dupuy16},~7 -- \citet{forveille05},~8 -- \citet{gizis00},~9 -- \citet{harding13},~10 -- \citet{jenkins09},~11 -- \citet{lepine09},~12 -- \citet{lodieu05},~13 -- \citet{mclean12},~14 -- \citet{osten09},~15 -- \citet{reid03},~16 -- \citet{schlieder14},~17 -- \citet{williams15},~18 -- \citet{wolszczan14}. $\blacklozenge$ -- The listed mass is the total mass of the binary. $\star$ -- The period for BRI0021-0214 is taken from \citet{harding13}; the length of the observations reported in the paper is shorter than the rotational period of the dwarf, so it can not be well determined.}

\centering
\label{parameters}
\begin{tabular}{l l c c c c c c c}

\hline
\hline
2MASS number		&	Other names	&	Sp. Type	&	M 								&	Period 				&	Distance			&	Flux density		& $\nu$	&	References\\		
					&				&			&	(M$_{\odot}$)						&	(hours)				&	(pc)				&	 ($\mu$Jy)		& (GHz)	&		             \\
\hline

J00242463--0158201	& BRI B0021-0214	&	M9.5	&	$<$ 0.06							&	($\sim$5)$^{\star}$	&	11.45 $\pm$ 0.55	&	83 $\pm$ 18		&	8.46		&	12,3,9,4,2\\
J00275592+2219328	& LP 349-25		&	M8+M9	&	0.121$\pm$ 0.009$^{\blacklozenge}$	&	1.86 $\pm$ 0.02		&	13.10 $\pm$ 0.28	&	365 $\pm$ 16		&	8.46		&	7,9,9,9,14\\
J00361617+1821104 	& --				&	L3.5		&	0.06 -- 0.074						&	3.0 $\pm$ 0.7			&	8.75 $\pm$ 0.06	&	259 $\pm$ 19		&	4.86		&	8,9,9,4,1\\
J13142039+1320011	& NLTT 33370		&	M7		& 	0.176 $\pm$ 0.002$^{\blacklozenge}$	&	3.7859 $\pm$ 0.0001	&	16.39 $\pm$ 0.75	&	1156 $\pm$15	&	4.89		&	11,6,17,11,13\\
J15010818+2250020	& TVLM513-46546	&	M9		&	0.06 -- 0.08						&	1.95958 $\pm$ 0.00005	&	10.59 $\pm$ 0.06	&	318 $\pm$ 9		&	8.46		&	10,9,18,4,5\\
J18353790+3259545	& LSR J1835+3259	&	M8.5	&	$<$ 0.083 						&	2.845$ \pm$ 0.003		&	5.67 $\pm$  0.02	&	525 $\pm$ 15		&	8.46		&	15,9,9,14,2\\
 
\hline

\end{tabular}
\end{minipage}
\end{table*}

\begin{figure*}
\includegraphics[width=17cm]{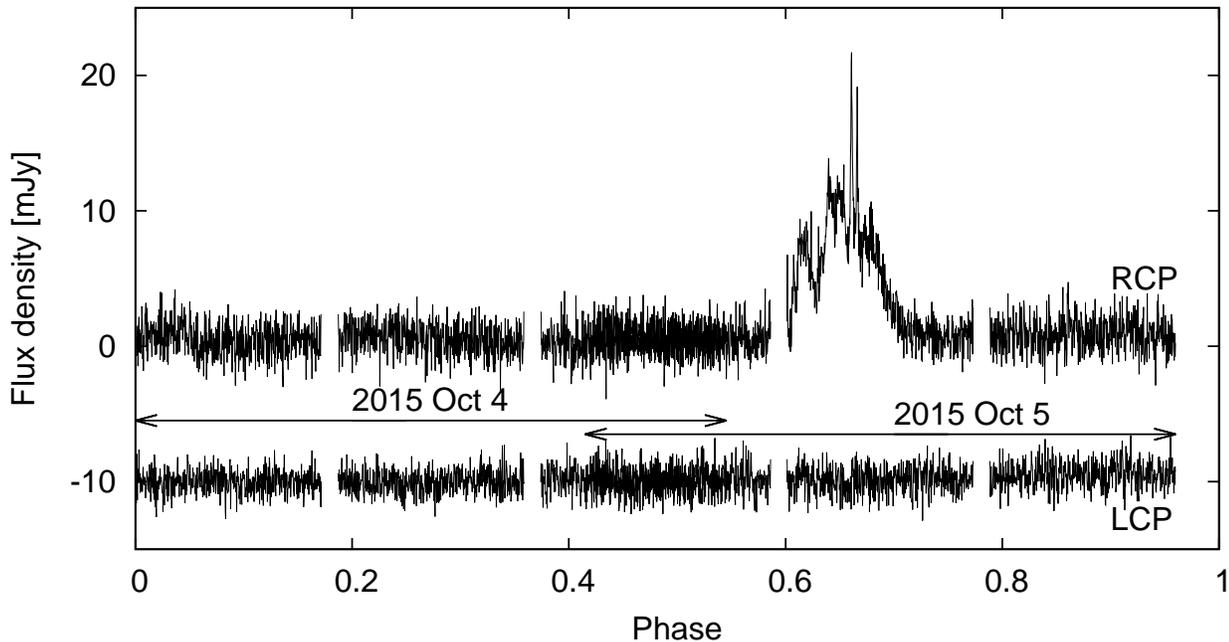} 
\vspace*{-1cm}
\caption[width=\textwidth]{Phased light curves for LSR 1835 from 3500 to 3756 MHz from 4 and 5 October 2014. The two observations have phase overlap from 0.4 to 0.55 and are plotted simultaneously. The LCP light curve is shifted down by 10mJy. Because the observations here cover almost one full rotational period, the x axis is phase, not time. \citet{hallinan15} show that for LSR 1835 events like that shown above are periodic, and occur every rotational period.\

}
\label{lsr-phased} 
\end{figure*}

\section{Target selection}\
\label{targets}

In selecting our targets, we have used the work of \citet{mclean12} and \citet{williams14}. All the sources we have observed have a range in spectral type between M7-- L3.5 and have been previously detected at radio wavelengths (see Table~\ref{parameters} for details).

\subsection{BRI B0021-0214}

Discovered by \citet{irwin91}, this M9.5 dwarf initially appeared to be inactive, showing very little H$\alpha$ and no X-ray emission \citep{basri95, reid99, neuhauser99, reiners10, cook14}. Detection in H$\alpha$ was reported by \citet{reid99} -- a small flare with L$_{H\alpha}$/L$_{bol}$ = -6.3, and later by \citet{berger10} with L$_{H\alpha}$/L$_{bol}$ = -6.05. \citet{martin01} reported optical periodicity ($\sim$4.8 and $\sim$20 hr), perhaps due to dust clouds in the atmosphere \citep{chabrier00}. \citet{harding13} also detected optical variability with a period of approximately 5 hours, the length of the observations reported in the paper is shorter than the rotational period of the dwarf, so the period can not be well determined. \citet{neuhauser99} reported a non-detection in X-rays with an upper limit of L$_X$/L$_{bol}$ $\leqslant$ 2.1 $\times$ 10$^{-5}$. \citet{berger10} found variable H$\alpha$ emission on a $\sim$0.5 - 2 hour time scale, but no radio emission. BRI B0021-0214 (hereafter BRI B0021) was previously detected in the radio at 8.46 GHz by \citet{berger02}, with a quiescent flux density of 83$\pm$18 $\mu$Jy. \citet{berger02} also reported a small flare with a flux density of 360 $\pm$ 70 $\mu$Jy (again at 8.46 GHz). BRI B0021 was later observed in the radio by \citet{berger10} and \citet{mclean12} with no detection.

\subsection{LP 349-25}

LP 349-25 was first recognised as a nearby late-type star by \citet{gizis00} based on its colour, parallax and proper motions \citep{reid03,salim03,gatewood05,lepine09}. It was classified as a M7.5V+M8.5V or a M8V+M9V binary by \citet{forveille05}. \citet{harding13} reported an optical rotational period of 1.86$\pm$0.02 hr. It was first detected in the radio by \citet{phan-bao07} with total flux density of the two (unresolved) sources of 365$\pm$16 $\mu$Jy and an upper limit of $\approx$13\% for a circularly polarised signal. \citet{osten09} detected radio emission from LP349-25, which was steady on both short and long time scale (minutes to months). The constant flux density from the source from epochs seven months apart suggested the presence of a long-lived magnetic structure giving rise to the emission. \citet{osten09} also reported the lack of rotational modulation of the radio light curve as significant and unique among the ultracool dwarfs.\

\subsection{2MASS J00361617+1821104 }

The L3.5 dwarf 2MASS J00361617+1821104 \citep{reid99} (hereafter 2M0036) was first detected in the radio by \citet{berger02} with a flux density of 135 to 330 $\mu$Jy and a low degree of circular polarisation (f$_c$ $\approx$ 13 - 35\% LCP). An approximately 20 minute burst of radio emission with a flux density of 720$\mu$Jy (f$_c$ $\approx$ 62\% LCP) was reported by \citet{berger02}. The detected radio emission was strongly circularly polarised and variable with an average flux density of 259 $\pm$ 19 $\mu$Jy and f$_c$ $\approx$ 73 $\pm$ 8\% LCP at 4.9GHz, and 134 $\pm$ 16 $\mu$Jy and f$_c$ $\approx$60 $\pm$15\% LCP at 8.5GHz. Simultaneous X-ray, H$\alpha$ and radio observations by \citet{berger05}, show no detection in either X-ray or H$\alpha$ with upper limits of L$_X$/L$_{bol}$ $\lesssim$ 2$\times$10$^{-5}$ and L$_{H\alpha}$/L$_{bol}$ $\lesssim$  2$\times$10$^{-7}$. The ratio of radio to X-ray luminosity is more than 4 orders of magnitude in excess of the G{\"u}del-Benz relation \citep{guedel93, benz94}. Both \citet{berger05} and \citet{harding13} reported an optical period of $\sim$3 hours.\

\subsection{NLTT 33370}

NLTT 33370 is a high proper motion star \citep{luyten79}, with $\mu$RA = -0.244'' yr$^{-1}$, and $\mu$Dec = -0.186'' yr$^{-1}$ \citep{schlieder14}. It was later reported as a young (80.8 $\pm$ 2.5 Myr \citep{dupuy16}) tight binary, with a distance between the companions of only 2.1 AU \citep{schlieder14}. The total mass of the system is 184.5 $\pm$ 1.6 M$_J$ (or 0.176 $\pm$ 0.002 M$_{\odot}$), and the two components have masses of 92.8 $\pm$ 0.6 and 91.7 $\pm$ 1.0 M$_J$ \citep{dupuy16}. The distance to NLTT 33370 is 17.249 $\pm$ 0.013 pc \citep{forbrich16}.\

NLTT 33370 is the brightest known radio UCD with stable emission \citep{becker95, mclean11, mclean12,dupuy16,forbrich16} varying sinusoidally with period of 3.89 $\pm$ 0.05 hours and total flux density from the unresolved binary of $\approx$ 1mJy at 4.9 GHz and 0.8 mJy at 22 GHz. As a result of simultaneous multi-wavelength observations, \citet{williams15} reported extreme magnetic activity as detected in X-rays, UV, broadband optical, H$\alpha$ and the radio, with variability observed in all wavelengths. Flaring  events were observed in all but the broadband optical observations. It was found that NLTT 33370 had two detectable periods of 3.7859 $\pm$ 0.0001 hours and 3.7130 $\pm$ 0.0002 hours, perhaps due to migration of spots or slightly different rotation periods of the two components of the binary. A similar periodicity was seen in the radio. The radio emission had three components - periodically modulated emission, with a periodicity similar to the optical, and $\sim$15\% RCP emission; frequent rapid 100\% LCP bursts (suggesting magnetic field strength of $\approx$2.1 kG) at all rotational phases (suggesting nearly continuous reconnection); and a variable unpolarised component.\

\subsection{TVLM 513-46546}

The radio emission from TVLM 513 was first reported to be variable by \citet{berger02} and later on from \citet{hallinan06}, with a period of about 2 hours. This was confirmed as the rotational period of the dwarf by \citet{lane07}, later refined to 1.95958 $\pm$ 0.00005 hours from optical observations by \citet{harding13} and \citet{wolszczan14}. Highly polarised radio bursts were reported by \citet{hallinan07, berger08} and \citet{doyle10}. \citet{doyle10} showed the data contained a series of consecutive periodic bursts with stable shape of the light curve, but changing shape/structure of the bursts. \citet{wolszczan14} reported stable optical and radio emission for 7 and 5 years respectively, that produce the same period with precision up to $\sim$20ms. \citet{berger08}, \citet{doyle10} and \citet{wolszczan14} report a correlation between the H$\alpha$ peaks and radio flares with approximately 0.4 phase offset, suggesting that the optical and radio variability originates from a large-scale dipolar magnetic field that is stable for at least a decade. \citet{wolszczan14} also reported that the period between bursts may be  gradually shortening, possibly due to migration of spots towards the equator and differential rotation.\

\subsection{LSR J1835+3259}

LSR J1835+3259 (hereafter LSR 1835) is another well-known active UCD. First recognised as a late-type dwarf by \citet{reid03}, it has high proper motions ($\mu$RA = -0.040'' yr$^{-1}$, and $\mu$Dec = -0.759'' yr$^{-1}$ \citep{schmidt07}), and an optical period of 2.845(3) hours \citep{harding13}. \citet{berger08} obtained simultaneous multi-wavelength observations of LSR 1835 in X-rays, UV, optical and the radio. The source was not detected neither in X-rays, nor in the UV. The reported upper limit for the X-rays of F$_X$ $<$ 8.4 $\times$ 10$^{-16}$ erg cm$^{-2}$ s$^{-1}$ (or L$_X$/L$_{bol}$ $<$ 10$^{-5.7}$) is one of the faintest X-ray limits for an UCD. \citet{berger08} also reported highly variable H$\alpha$ emission and nearly constant radio emission. \citet{hallinan08, hallinan15} reported persistent 100\% circularly polarised bursts of radio emission with a period similar to the optical, and associated magnetic field strength of $\sim$3 kG. A similar estimation was obtained by \citet{berdyugina15} by using spectropolarimetric methods: they estimated the average magnetic field strength at the surface of LSR 1835 as $\sim$2.5 kG. By assuming a dipole-like magnetic geometry, one obtains magnetic field strength of $\sim$3.6 kG at the magnetic poles.\\

\begin{table*}
\begin{minipage}{180mm}
\caption{Observed targets, starting date of the observations, scheduling block names, antennae configuration of the VLA, flux and phase calibrators, and notes. All observation blocks last for 2 hours including overheads.}
\centering
\label{observations}
\begin{tabular}{l c r c c c c}

\hline
\hline	
Target		&	Start of			&	Scheduling	& 	Antennae 			&	Flux			&	Phase		&	Notes			\\
			&	observations		&	block name	&	configuration			&	calibrator	&	calibrator	&					\\
\hline

BRI B0021	&	2014-10-06T03:11	&	BRI0021 (1)	&	DnC					&	3C48		&	J0022+0014	&	upper limit 		\\
LP 349-25	& 	2014-10-06T08:40	&	LP349-25 (1)	&	DnC					&	3C48		&	J0029+3456	&	detected, no flares	\\
			&	2014-10-07T03:27	&	LP349-25 (2)	&	DnC $\rightarrow$ C	&	3C48		&	J0029+3456	&	detected, no flares	\\
			&	2014-10-08T01:48	&	LP349-25 (3)	&	DnC $\rightarrow$ C	&	3C48		&	J0029+3456	&	detected, no flares	\\
			&	2015-01-28T01:12	&	LP349-25 (4)	&	CnB $\rightarrow$ B	&	3C48		&	J0029+3456	&	detected, no flares	\\
2M0036 		& 	2014-10-07T08:46	&	2M0036 (1)	&	DnC $\rightarrow$ C	&	3C48		&	J0122+2502	&	detected, no flares	\\
			&	2014-10-08T03:48	&	2M0036 (2)	&	DnC $\rightarrow$ C	&	3C48		&	J0122+2502	&	detected, no flares	\\
NLTT 33370	& 	2015-02-05T12:09	&	NLTT33370 (1)	&	CnB $\rightarrow$ B	&	3C286		&	J1347+1217	&	detected, no flares	\\
TVLM 513	&	2015-02-06T15:04	&	TVLM513 (1)	&	CnB $\rightarrow$ B	&	3C286		&	J1513+2338	&	detected, no flares	\\
LSR 1835		&	2014-10-04T03:19	&	LSR1835 (1)	&	DnC					&	3C48		&	J1924+3329	&	detected, no flares	\\
			&	2014-10-05T03:15	&	LSR1835 (2)	&	DnC					&	3C48		&	J1924+3329	&	100\% RCP flare	\\

\hline

\end{tabular}
\end{minipage}
\end{table*}

%-------------------------------------

\section{Instruments, observations and data reduction} \label{observations}\

The observations were carried with the Karl Jansky Very Large Array (VLA) in semester 2014B (Project ID: 14B-015) simultaneously in S (2--4 GHz) and C (4--8 GHz) band in two sub-arrays, using 14 and 13 antennae respectively. All the observations were done in two-hour blocks (with approximately 1.5 hours on-source time) with time and frequency resolutions of 3 s and 1 MHz respectively. More detailed information about the observations is presented in Table~\ref{observations}. \

The data were reduced using the Common Astronomy Software Applications package (CASA) versions 4.2.0 and 4.2.2. The calibration was done with the EVLA pipeline version 1.3.1, and for comparison it was also done manually for three datasets. The results were consistent within the error bars. Polarisation calibration was not performed for those short time observations. The non-zero instrumental polarisation in this case is around or less than 1 per cent, which is negligible in the case of 100 per cent burst events.
Radio frequency interference (RFI) was flagged both manually and automatically using the AOFlagger tool \citep{offringa10, offringa12}. Additional flagging was performed where needed. Radio bursts were distinguished from RFI based on the signal in Stokes Q and U (since RFI is strongly linearly polarised, and most of the bursts from these UCDs are mainly circularly polarised) with the affected channels flagged in all polarisations. Additional checks for RFI were performed with the Multichannel Image Reconstruction, Image Analysis and Display package (MIRIAD) \citep{sault95}. \

To develop a map and model of the radio emission from background sources in the field, the data were imaged using CASA's multi-frequency synthesis \citep{sault94} and CASA's multi-frequency CLEAN algorithm. We created maps of 3000$\times$3000 pixels, each 1$''$$\times$1$''$, and after subtracting the background sources, new images ($\sim$200$''$$\times$200$''$) were created of just the source. All the detected sources exhibit radio emission which is consistent with zero linear polarisation. Photometry was extracted from the visibility-domain data in Stokes I and V for all the sources. The results are presented in Table~\ref{results}, where the error bars for the measurements are at 1 $\sigma$, and the upper limits are at 3 $\sigma$ level. \\

%-------------------------------------

\section{Results}\

\begin{figure*}
\begin{center}
\includegraphics{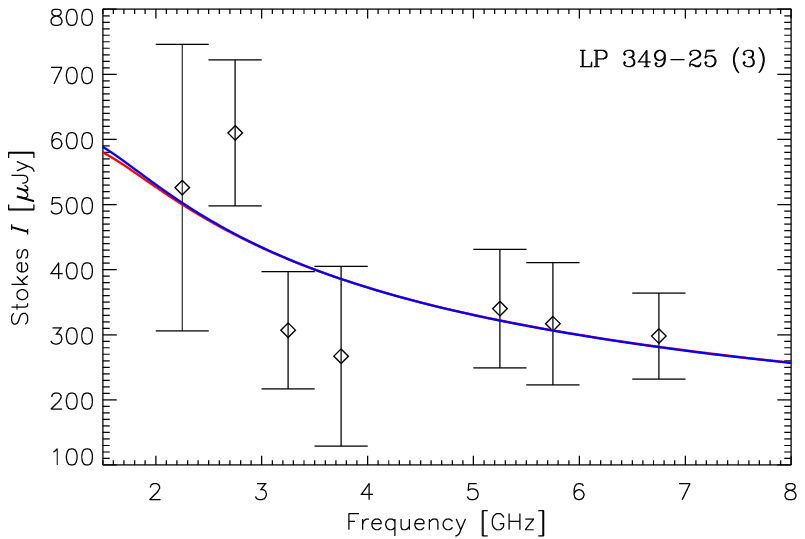}
\includegraphics{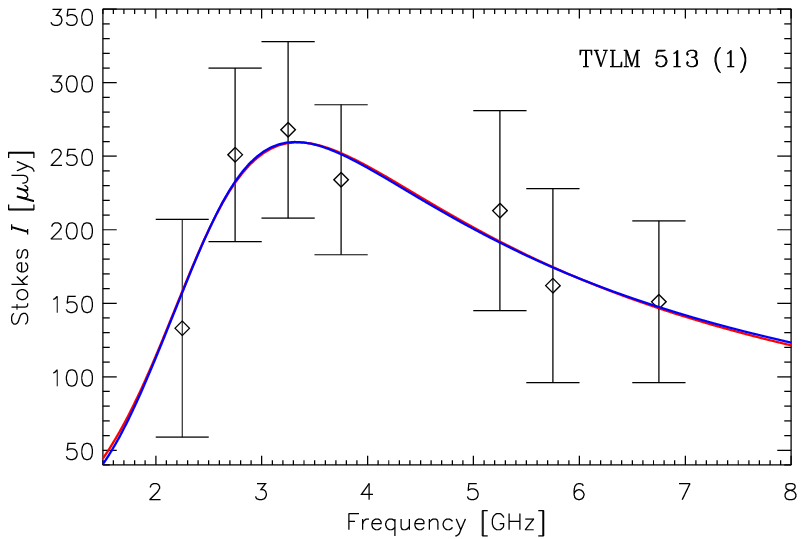}
\includegraphics{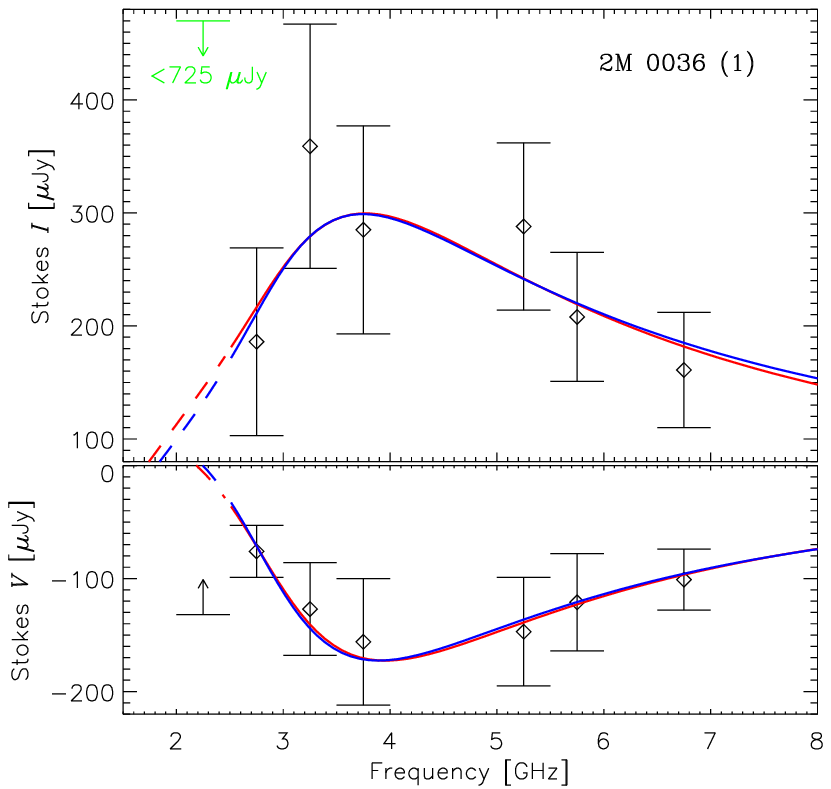}
\includegraphics{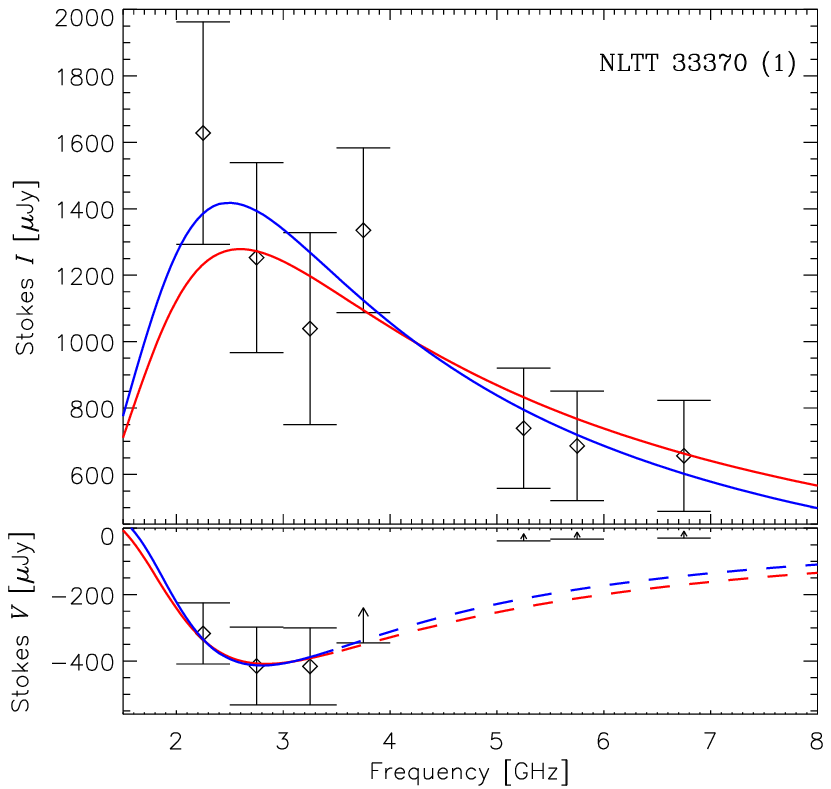}
\includegraphics{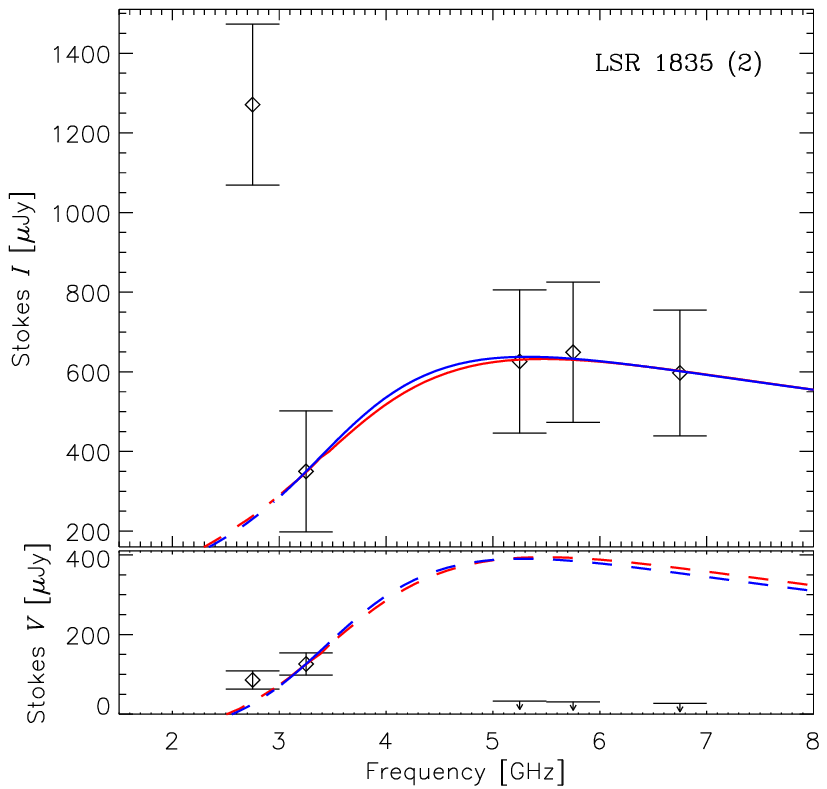}
\end{center}
\caption{Selected radio spectra of LP 349-25, TVLM 513, 2M 0036, NLTT 33370, and LSR 1835. For clarity, if an object was observed several times, only one spectrum is shown, since the spectrum shape remains nearly the same. The circular polarisation (Stokes $V$) is shown only if it was detected. The red and blue curves represent the simulated gyrosynchrotron spectra for the parameter sets corresponding to the surface magnetic strengths of $B_0=2000$ G (red) and 5000 G (blue) (see Section \protect\ref{Qsim} and Table \protect\ref{GSfits})  (The blue and red line overlap almost completely for most of the plots). Dashed curves indicate that the measured values or upper limits in the corresponding frequency ranges were not used for spectral fitting.}
\label{SED}
\end{figure*}

\begin{figure*}
\centering
$
\begin{array}{cc}
\hspace{0cm}
\includegraphics[height=9cm]{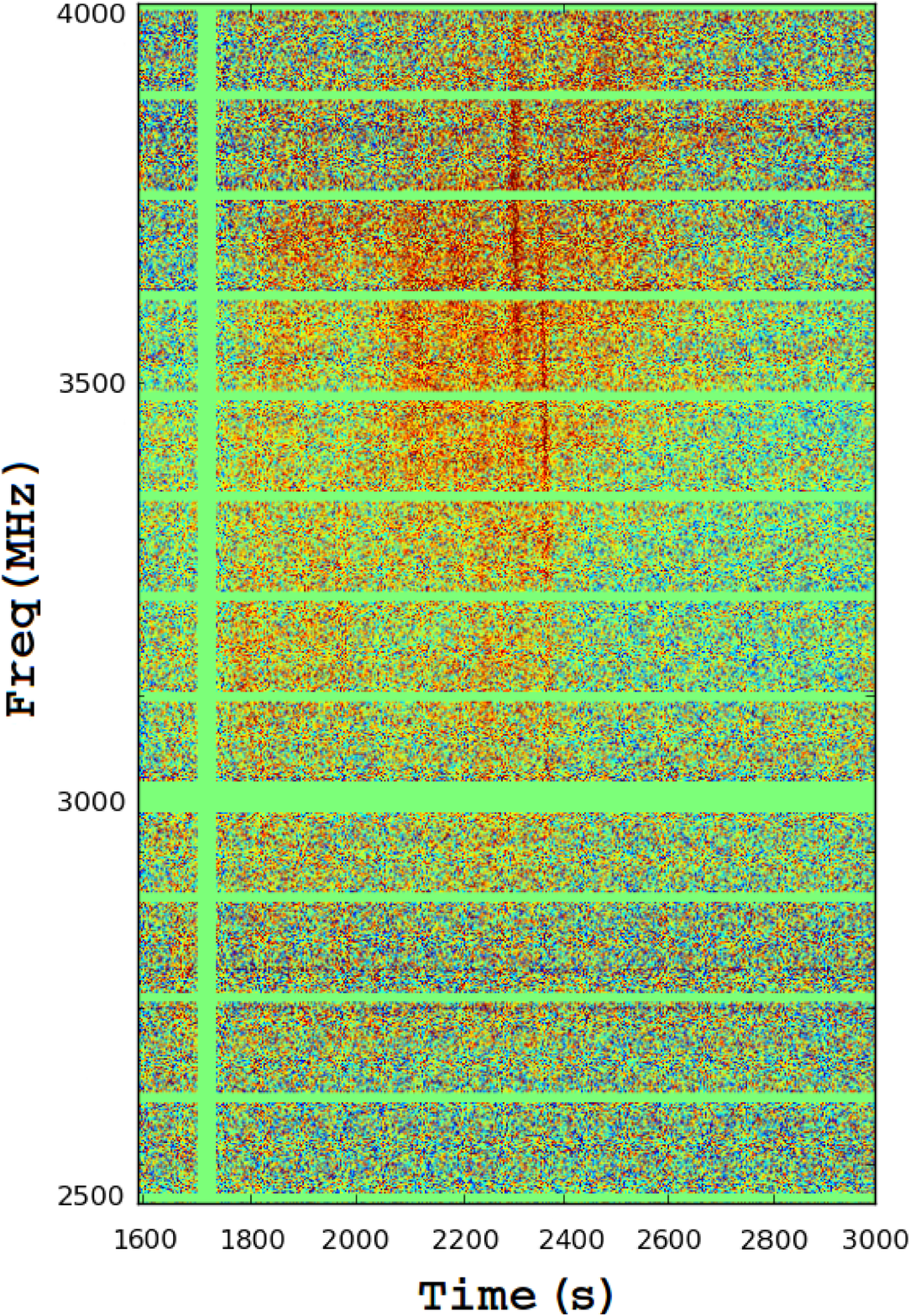} 
\hspace{-1.5cm}
\includegraphics[height=9.2cm]{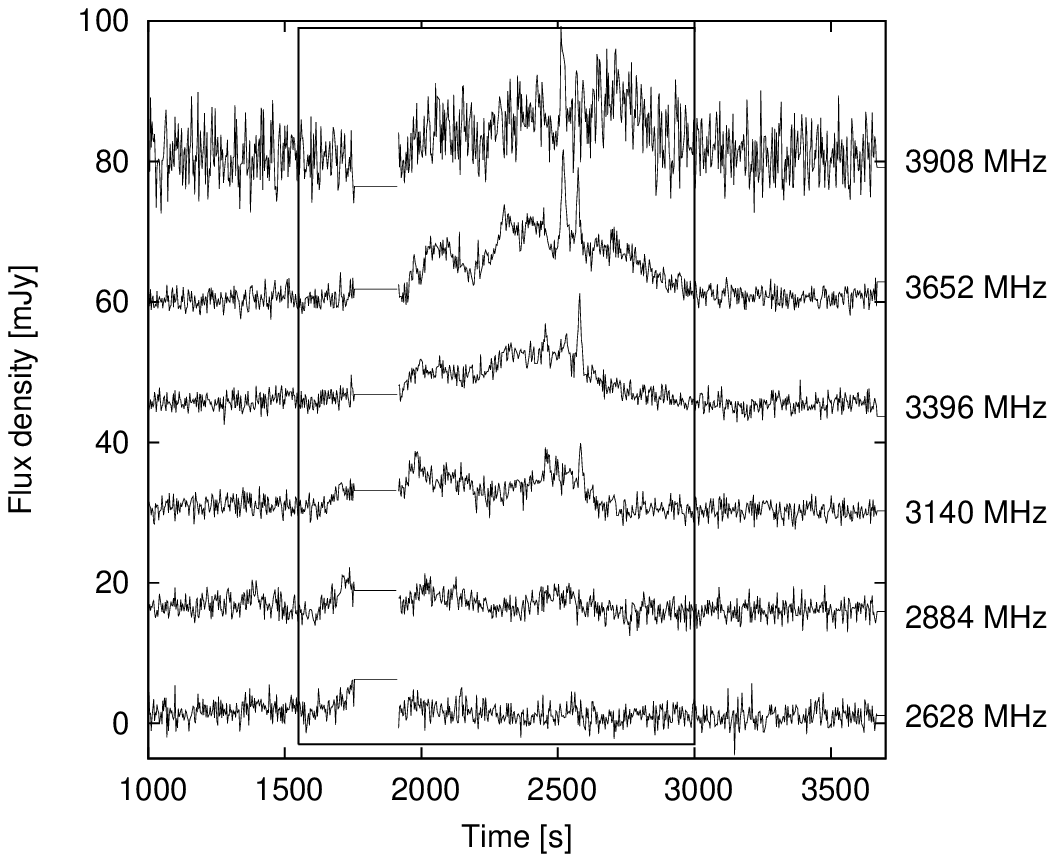} 
\end{array}$ \\
\caption[width=\textwidth]{Radio emission from LSR 1835. Observations were obtained on 2014-10-05T03:15, and again the time scale is in seconds after the beginning of the observational block. Left: dynamic spectrum (Stokes V, RCP) of the flare. Right: Light curves for the observation in S band (again in Stokes V, RCP) from 2.5 to 4.0 GHz (bottom to top) in 256 MHz bins. The labels on the right side show the central frequency for each bin. The rectangular on the right panel encloses the area of the corresponding dynamic spectrum from the left panel. The horizontal and vertical gaps in the dynamic spectrum are a result of the RFI flagging.}
\label{lsr-flare} 
\end{figure*}

From the six sources we have observed, significant radio emission was detected from five -- LP349-25, 2M0036, NLTT 33370, TVLM513 and LSR 1835. An additional upper limit was determined for the very quiet/inactive dwarf BRI B0021 of 46$\mu$Jy over the S band, and 21$\mu$Jy over the C band. Three types of radio emission features were observed -- unpolarised quiescent emission (only Stokes I), polarised quiescent emission (both Stokes I and V), and a flaring emission from the dwarf LSR 1835. For all S band observations, strong RFI is present in two spectral windows - 2.0--2.5 GHz and 3.5--4.0 GHz, and for C band - between 6.0 -- 6.5 GHz. Cleaning was performed both manually and with the AOFlagger software \citep{offringa10} in those windows, and additional manual flagging where needed (see Section~\ref{observations}). Upper limits and flux densities for all observations are listed in Table~\ref{results}, where the flux is listed when the remaining unflagged data covers more than 150 MHz. For LSR 1835, the quiescent flux density is measured after the flare was removed from the main observation.\

\begin{table*}
\begin{minipage}{180mm}
\caption{Flux density and upper limits in $\mu$Jy for all observations. Due to strong RFI in most of the frequency range 6 - 6.5 GHz, flux densities for those frequency windows are excluded. Missing points in this table were either due to excess RFI in that frequency range or a weak signal. Data for the spectral windows 2.0 -- 2.5 GHz and 3.5 -- 4.0 GHz is only presented for observations with more than 150 MHz remaining in that window after RFI flagging.}
\centering
\label{results}
\begin{tabular}{l l r r r r r r r r r r r}

\hline
\hline
Target		&	SB Name				&	2.0 -- 2.5 GHz		&	2.5--3.0 GHz		&	3.0--3.5 GHz		&	3.5 -- 4.0 GHz		&	5.0--5.5 GHz	&	5.5--6.0 GHz	&	6.5--7.0 GHz    		\\
\hline

BRI B0021	&	BRI0021 (1) \textit{I}	&		-			&	$<$129			&	$<$47   			&		-			&	$<$28	       	&       $<$30		&  	 $<$88  		 	\\
			&	BRI0021 (1) \textit{V}	&		-			&	$<$99			&	$<$37   			&		-			&	$<$29	       	&       $<$29		&  	 $<$79  		 	\\

LP 349-25	&	LP349 (1) \textit{I}		&		-			&	409 $\pm$ 92    	&       347 $\pm$ 80    	&		-			&	325 $\pm$ 77	&       306 $\pm$ 73	&  	 247 $\pm$ 66		\\	
			&	LP349 (1) \textit{V}	&		-			&	$<$42   			&       $<$33	       		&		-			&	$<$32	      	&       $<$31	       	&  	 $<$26  		 	\\

%			&	LP349 (2) \textit{I}		&		-			&	$<$133  			&       126 $\pm$ 91    	&		-			&       94 $\pm$ 57	&       232 $\pm$ 60 &  	 354 $\pm$ 66		\\
%			&	LP349 (2) \textit{V}	&	 	-			&	$<$40   			&       $<$36	       		&		-			&       $<$33		&       $<$34	       	&  	 $<$29  			\\

			&	LP349 (3) \textit{I}		&	 526 $\pm$ 220	&	610 $\pm$ 112       &       307 $\pm$ 90    	&	267 $\pm$ 138	&       340 $\pm$ 91	&       317 $\pm$ 94 &  	 298 $\pm$ 66		\\
			&	LP349 (3) \textit{V}	&	 		-		&	$<$48  			&       $<$35	      		&		-			&       $<$37		&       $<$33	       	&  	 $<$32  		 	\\

			&	LP349 (4) \textit{I}		&		-			&	425 $\pm$ 99		&       321$\pm$ 91     	&		-			&       329 $\pm$ 81	&       298 $\pm$ 79 &  	 288 $\pm$ 69		 \\
			&	LP349 (4) \textit{V}	&		-			&	$<$46 			&       $<$38	      		&		-			&       $<$ 34		&       $<$ 34	       	&  	 $<$31  			 \\

2M0036		&	2M0036 (1) \textit{I}	&	$<$ 725			&	186 $\pm$ 83		&	359 $\pm$ 108	&	285 $\pm$ 92		&	288 $\pm$ 74	&	208$\pm$ 57	&  	161 $\pm$ 51		\\
			&	2M0036 (1) \textit{V}	&	$<$ 132			&	-76 $\pm$ 23		&	-127 $\pm$ 41	&	-156 $\pm$ 56	&	-147 $\pm$ 48&	-121	$\pm$ 43	&  	-101 $\pm$ 27	 \\

			&	2M0036 (2) \textit{I}	& 		-			&	138 $\pm$ 57     	&	604 $\pm$ 129	&		-			&       280 $\pm$ 67	&   291  $\pm$ 65    &  	 251 $\pm$46		 \\
			&	2M0036 (2) \textit{V}	&	 	-			&	-68 $\pm$ 33   	&	-158 $\pm$ 59    	&		-			&       -57 $\pm$ 16  &       $<$30	      	&  	 $<$26  			 \\

NLTT 33370	&	NLTT33370 (1) \textit{I}	&	1628 $\pm$ 335	&	1253 $\pm$  286     &	1039 $\pm$ 289	&	1335 $\pm$ 248	&	739 $\pm$181&	686 $\pm$ 165&  	656 $\pm$ 167	 \\
			&	NLTT33370 (1) \textit{V}&	-317 $\pm$ 92	&	-415 $\pm$ 117	&	-416 $\pm$ 116	&	$<$ 345			&	$<$38		&	$<$33   		&  	$<$30			 \\

TVLM 513	&	TVLM513 (1) \textit{I}	&	133 $\pm$ 74		&	251 $\pm$ 59		&	268 $\pm$ 60		&	234 $\pm$ 51	&	213 $\pm$ 68 	&	162 $\pm$ 66		&  	151 $\pm$ 55		 \\
			&	TVLM513 (1) \textit{V}	&	$<$ 53			&	$<$46			&	$<$45			&	$<$39			&	$<$41      		&	$<$ 44		& 	$<$35			 \\

LSR 1835		&	LSR1835 (1) \textit{I}	&		-			&	1132 $\pm$ 159	&	474$\pm$ 133	&		-			&     835 $\pm$ 206	&     759 $\pm$ 165	&  	806 $\pm$ 157 	\\
			&	LSR1835 (1) \textit{V}	&		-			&	122 $\pm$ 35		&	126 $\pm$ 29		&		-			&	$<$60		&	$<$56 		&  	$<$62			 \\
		
			&	LSR1835 (2) \textit{I}	&		-			&	1271 $\pm$ 202	&	350 $\pm$ 152	&		-			&	626$\pm$180	&	649$\pm$ 176&  597$\pm$ 158		 \\
			&	LSR1835 (2) \textit{V}	&		-			&	86 $\pm$ 23		&	126 $\pm$ 28		&		-			&	$<$33      		&	$<$31		&  	$<$27			 \\

\hline																

\end{tabular}
\end{minipage}
\end{table*}

\subsection{Quiescent emission}

Quiescent emission was observed from five sources. \

\textit{2M0036} had a strongly variable flux in different frequency regions with the intensity varying from $\approx$140 to $\approx$600 $\mu$Jy. The emission showed steady left-hand circular polarisation with levels of approximately 50 per cent for the first observation, decreasing significantly from about 50 to 20 per cent for the second. This source was observed twice, with very good agreement between the total flux in the two measurements (Table~\ref{results}). \

\textit{NLTT 33370} showed decreasing flux density from just over 1600 $\mu$Jy at 2.0 GHz to  $\sim$650 $\mu$Jy at 7GHz. The emission showed varying levels of left-hand circular polarisation of up to 40 per cent at low frequencies (below 3.5 GHz). \

\textit{TVLM 513} had a large variation in its flux density with frequency, changing from 150 to 270 $\mu$Jy (Fig.~\ref{SED}), with an upper limit for the circular polarisation of about 20 per cent.\ 

\textit{LP349-25} was observed four times, showing very good agreement between three of the measurements. The emission was unpolarised with an upper limit for the circular polarisation of approximately 10 per cent.  One of the observations was ignored due to strong RFI present for most of the time of the observation, and poor signal to noise ratio after flagging.\ 

\textit{LSR 1835} showed a strongly variable flux density with frequency -- varying from 350 to 1270 $\mu$Jy, and approximately stable flux level of circular polarisation of about 100 $\mu$Jy in frequencies below 4 GHz. LSR 1835 was observed twice in two consecutive days, with phase coverage of almost one full rotational period (see Fig. \ref{lsr-phased}), where phase 0 corresponds to the beginning of the first observation. During the second observation, a large burst occurred (see Section~\ref{flare-emission}). \

The spectral energy distributions for all sources are shown in Fig.~\ref{SED} with gyrosynchrotron fits added (see Section \ref{Qsim}). Since the quiescent spectra at different times (if an object was observed more than once) are very similar, we show only one spectrum for each object.\

\subsection{Flaring emission} \label{flare-emission}\

Flare-like emission was detected from only one source -- LSR 1835. The burst was detected during the observation session on 2014 Oct 5, at a  rotation phase of $\approx$0.625 (relative to the start of the first observation, see Fig. \ref{lsr-phased}). The event was detected in the S band (2--4 GHz) and in the RCP channel only, i.e. the emission was 100 per cent right-hand circularly polarised.\

\begin{figure}
\hspace{-1.6cm}
\includegraphics[width=12cm]{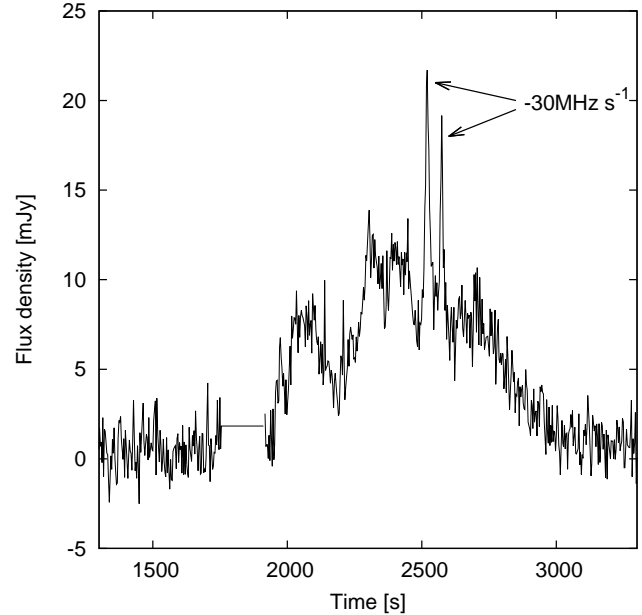} 
\caption{Different features within the burst of LSR 1835 as seen from 3500 to 3756 MHz -- a fainter long duration event with a positive frequency drift of 1 MHz s$^{-1}$ (approximately 1700 to 3100s), and bright short duration events with a negative frequency drift of about $-30$ MHz s$^{-1}$ (around 2500 -- 2600s).}
\label{lsr-features} 
\end{figure}

Figure~\ref{lsr-flare} shows the dynamic spectrum of the flare (left) and the evolution of the flare in different frequencies with corresponding light curves in 256 MHz bins in the $2.5-4.0$ GHz range (right). In addition, Figure \ref{lsr-features} shows an enlarged fragment of the light-curve ($3500-3756$ MHz), where this spectral window was selected due to strong signal (and thus easily distinguishable) in both components of the flare. It is noticeable that the flaring emission has a complicated spectrum consisting of several components. In particular, around 2500 -- 2600s  there are two easily identifiable intense narrowband bursts. Those features have flux density exceeding 20 mJy, and a fast negative frequency drift of about $-30$ MHz $\textrm{s}^{-1}$. The bursts have the same frequency drift rate (Fig.~\ref{lsr-flare}). They are superimposed on a fainter broadband emission feature with a total duration of about 20 minutes ( $\sim$1700s to 3100s), bandwidth of about 1 GHz, flux density of up to 10-13 mJy, and a slow positive frequency drift of about 1 MHz $\textrm{s}^{-1}$.

In earlier observations of \citet{hallinan15}, LSR 1835 was found to produce periodic radio bursts with a period coinciding with its rotation period. However, since our observations covered only about one stellar rotation (with a very small phase overlap, 0.40 -- 0.55, of the observations on two consecutive days), we cannot tell whether the detected radio flare was a single or a periodic event. The observed dynamic spectrum with several narrowband and broadband features implies a complicated structure of the emission source (see Section \ref{Fsim}).\

%--------------------------------------

\section{Modelling}\label{modelling}
\subsection{Quiescent radio emission}\label{Qsim}
\subsubsection{Source model}\label{hommod}\

The broadband quiescent (i.e., constant or slowly varying) radio emission of ultracool dwarfs is usually interpreted as incoherent gyrosynchrotron emission of energetic electrons in a relatively weak magnetic field \citep[][etc.]{berger01, berger02, osten06b, ravi11}. The gyrosynchrotron emission spectra are characterized by the optically thick part (with a positive slope) at low frequencies and the optically thin part (with a negative slope) at higher frequencies. As can be seen in Figure \ref{SED}, our observations agree qualitatively with the gyrosynchrotron model; the only exception is LSR 1835 which will be discussed in Section~\ref{Fsim}. The observed circular polarisation degrees of the quiescent emission (from zero to approximately 50\%) are also consistent with the predictions for the gyrosynchrotron mechanism \citep{dulk85}.\

To estimate the parameters of the magnetic field and energetic electrons in the magnetospheres of ultracool dwarfs, we have performed numerical simulations of their gyrosynchrotron emission. Since only a limited number of data points is available, we use the simplest (i.e., with the smallest number of parameters) source model: a homogeneous emission source with the depth (along the line-of-sight) of $L$ and the visible area of $L^2$. The uniform magnetic field is characterized by its strength $B$ and viewing angle $\theta$ (relative to the line-of-sight). The energetic electrons are assumed to have an isotropic power-law spectrum with the spectral index $\delta$ in the energy range from $E_{\min}=10$ keV to $E_{\max}=100$ MeV; the chosen energy range is consistent with earlier simulations \citep{osten06b, ravi11, lynch16} and agrees also with the estimations for the electrons in the Jovian radiation belts \citep[][see details below]{santos08}. Thus, the emission spectrum (Stokes $I$ and $V$) depends on five parameters: source size $L$, magnetic field strength $B$ and viewing angle $\theta$, the total concentration of energetic electrons $n_{\mathrm{b}}$ and their spectral index $\delta$. The emission spectra have been computed using a fast gyrosynchrotron code \citep{fleishman10}.\

As can be seen in Table \ref{results}, the circular polarisation of the radio emission from ultracool dwarfs is highly variable: for some objects, the polarisation degree reached 50\%, while for others (LP 349-25 and TVLM 513) the Stokes $V$ flux density was below the threshold of detectability. We believe that the non-detectable polarisation might be caused actually by the source inhomogeneity: e.g., in a dipolar magnetosphere, if the magnetic dipole is nearly perpendicular to the line-of-sight, the Stokes $V$ fluxes with opposite signs from opposite magnetic hemispheres should compensate each other resulting in low total polarisation. This effect cannot be reproduced by our simplified source model. Therefore, for the dwarfs LP 349-25 and TVLM 513 we do not consider the emission polarisation but only the intensity; in the calculations for these objects the viewing angle is fixed and set to $\theta=80^{\circ}$. A similar effect may account for non-detectable circular polarisation of high-frequency emission from NLTT 33370 and LSR 1835 (see the discussion below in Section \ref{mrdiscussion}); again, in the corresponding frequency ranges we consider only the emission intensity. Note also that we consider here the emission flux densities accumulated over the time intervals comparable with the rotation periods of the dwarfs, therefore the possible emission variations caused by the source rotation are expected to be smoothed out which may result, e.g., in further decrease of the observed polarisation degree.

\subsubsection{Markov chain Monte Carlo analysis}\

To find the model parameters that provide the best agreement with the observations, we have used both the simple least-squares fitting and the Markov chain Monte Carlo (MCMC) analysis \citep[see, e.g.,][]{Gregory05}. In particular, the adopted likelihood function $p$ is given by

\begin{equation}\label{MCMCprobability}
p=Ab^{N/2}\exp(-b\chi^2),\qquad
\chi^2=\sum\limits_{i=1}^N\frac{(S_i^{\mathrm{mod}}-S_i^{\mathrm{obs}})^2}{2\sigma_i^2},
\end{equation}

where $S^{\mathrm{obs}}_i$ and $S^{\mathrm{mod}}_i$ are the observed and computed radio emission flux densities (including both Stokes $I$ and $V$), respectively, $\sigma_i$ are the respective measurement errors, and $A$ is the probability normalization factor. The number of data points $N$ can be different for different datasets; as mentioned above, the Stokes $V$ values have been used in the model fitting procedure only if the polarised signal was reliably detected. The parameter $b$ is the noise scale parameter that is introduced to account for possible overestimation of the measurement errors \citep{Gregory05}. As the priors for all model parameters we use uniform distributions in the ranges broad enough to cover all feasible parameter sets; since the energetic electrons concentration can vary by several orders of magnitude, we use $\log n_{\mathrm{b}}$ instead of $n_{\mathrm{b}}$ as the model parameter. The posterior distributions have been produced using the Metropolis-Hastings algorithm with the parameters adjusted to provide the acceptance rate of about 0.25 \citep[see][]{Gregory05}. The produced MCMC chains in our simulations consisted of $\sim 10^7$ steps which has been found to be sufficient to achieve stable posterior distributions. At the same time, the best least-squares fit for each dataset evidently corresponds to the global minimum of $\chi^2$ and is independent on the noise scale parameter $b$.\

\begin{figure*}
\begin{center}
\includegraphics{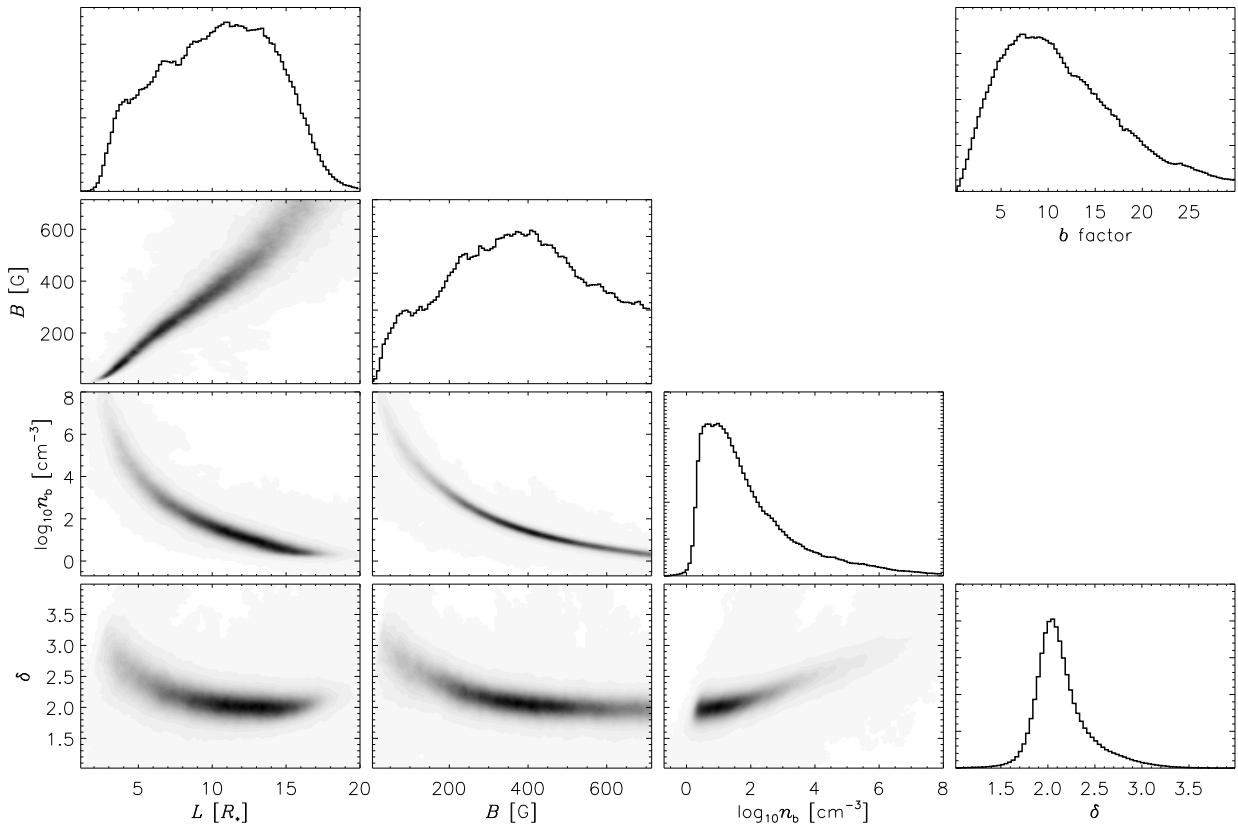}
\end{center}
\caption{Results of the MCMC analysis of gyrosynchrotron emission model for the dwarf TVLM 513: posterior probability distributions. Greyscale plots show the joint posterior probability distributions for different combinations of parameters (darker areas correspond to higher probability), while marginal posterior distributions are shown as histograms. The marginal posterior distribution for the noise scale parameter $b$ is shown in the upper right corner.}
\label{MCMC1}
\end{figure*}

Figure \ref{MCMC1} demonstrates, as a typical example, the posterior probability distributions computed for the dwarf TVLM 513 (note that the viewing angle $\theta$ for this object is assumed to be fixed at $\theta=80^{\circ}$). The upper limit of the magnetic field strength $B$ is 715 G, because this value corresponds to the electron cyclotron frequency of 2 GHz that is the lower boundary of the observed radio spectrum. One can see that the existing data constrain the model parameters only partially. Most importantly (from the physical point of view), there are virtually no constraints on the characteristic source size $L$: this parameter can vary in a broad range from ${\approx 2R_*}$ to ${\gtrsim 10R_*}$, where $R_*$ is the radius of the dwarf (which is assumed to equal 70\,000 km for all objects in this work). The typical values of the magnetic field strength $B$ and energetic electrons concentration $n_{\mathrm{b}}$ also vary in broad ranges as functions of $L$; the electron spectral index $\delta$ is better defined but still variable. As can be seen in Figure \ref{SED}, different parameter sets can indeed provide equally good fits to the observed data (see the red and blue curves overplotted on the observed spectra). Similar results were obtained earlier by other authors in application to a number of ultracool dwarfs \citep{osten06b, ravi11, lynch15, lynch16}. We note that the mentioned degeneracy between different model parameters is not removed even if the polarisation measurements are available (like in the case of 2M 0036), in contrast to the suggestions of \citet{lynch15}.\

\begin{figure*}
\begin{center}
\includegraphics{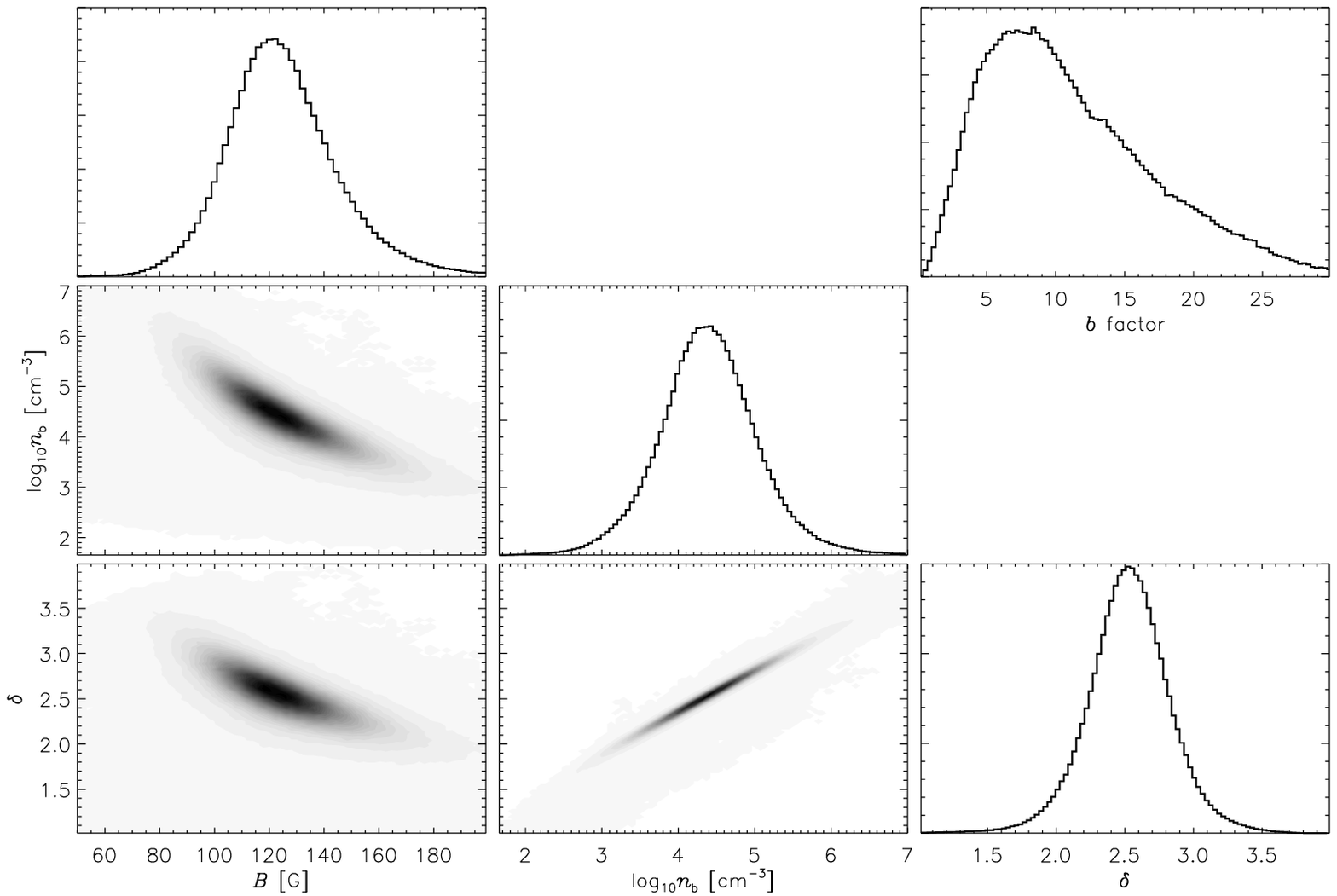}
\end{center}
\caption{Results of the MCMC analysis of gyrosynchrotron emission model for the dwarf TVLM 513 (same as in Figure \protect\ref{MCMC1}) in the case when the characteristic source size $L$ is fixed (${L=4.60R_*}$).}
\label{MCMC2}
\end{figure*}

On the other hand, we have found that for a fixed source size $L$ all other parameters can be effectively constrained. Figure \ref{MCMC2} shows the posterior probability distributions computed for TVLM 513 in the case when the source size is fixed and set arbitrarily to ${L=4.60R_*}$. One can see that now the probability distributions of the magnetic field strength $B$, energetic electrons concentration $n_{\mathrm{b}}$ and electron spectral index $\delta$ demonstrate well-defined narrow peaks corresponding to the best-fit set of the model parameters (which corresponds to the global minimum of $\chi^2$ as well).\

\subsubsection{Restrictions on the model parameters}\

For the above-described reasons, instead of trying to find a unique best-fit set of the emission source parameters, we now explore relations between these parameters. To put additional physical constraints on the source size and magnetic field strength, we consider an analogy with the incoherent radio emission of Jupiter. The Jovian decimetric radiation is produced by high-energy electrons in the Jovian radiation belts due to the gyrosynchrotron mechanism \citep{carr83, zarka00}; hence we suggest that the quiescent radio emission of ultracool dwarfs may be produced in a similar way, i.e., it may be considered as an ``up-scaled'' version of the Jovian decimetric radiation. According to imaging radio observations \citep[see, e.g.,][]{bolton02, santos09, santos14, girard16}, the source of Jovian decimetric radiation has the shape of a torus located in the equatorial plane, with the major radius of about $R_{\mathrm{c}}\simeq 1.5R_{\mathrm{J}}$ and the minor radius of about $r\simeq 0.5R_{\mathrm{J}}$ (i.e., $r\simeq R_{\mathrm{c}}-R_{\mathrm{J}}$). By assuming a similar source geometry for ultracool dwarfs, we estimate the volume of the emission source as

\begin{equation}\label{Vtor}
V=2\pi^2R_{\mathrm{c}}r^2\simeq 2\pi^2R_{\mathrm{c}}(R_{\mathrm{c}}-R_*)^2.
\end{equation}

For a dipole-like magnetic field, the magnetic field strength $B$ at the minor axis of this toroidal emission source (i.e., at the distance of $R_{\mathrm{c}}$ from the dwarf centre in the equatorial plane) is given by

\begin{equation}\label{Btor}
B=\frac{B_0}{2}\left(\frac{R_{\mathrm{c}}}{R_*}\right)^{-3},
\end{equation}

where $B_0$ is the maximum surface magnetic field strength (at the magnetic pole); the value given by this Equation can be considered as an average field strength in the toroidal source. We assume now that the toroidal emission source can be represented approximately by the homogeneous emission source described in Section \ref{hommod}, provided that the homogeneous source volume $V=L^3$ equals the toroidal source volume given by Equation (\ref{Vtor}) and the magnetic field strength $B$ equals the characteristic value given by Equation (\ref{Btor}); the viewing angle $\theta$ in the toroidal source model corresponds approximately to the angle between the magnetic dipole and the line-of-sight. Thus, for given source size $L$ and magnetic field strength $B$, we can estimate the corresponding magnetic field strength at the surface level $B_0$. Note that reducing a radiation belt-associated emission source to a homogeneous one is obviously a very rough approximation; we use it only to estimate the feasible variation ranges for the basic source parameters.

\begin{table*}
\caption{Results of the MCMC analysis of gyrosynchrotron emission model for several chosen datasets. The table lists the chosen characteristic source size $L$ (fixed during simulations) and the corresponding most probable values (within 75\% confidence limits) of the magnetic field strength $B$, the viewing angle $\theta$, the concentration of energetic electrons $n_{\mathrm{b}}$, the electron spectral index $\delta$, and the noise scale parameter $b$. For reference, the corresponding best-fit values obtained by minimizing $\chi^2$ with the gradient-expansion method are given in brackets (they are independent of $b$). The chosen source sizes $L$ correspond to the equivalent surface magnetic field of $B_0=3500$ G (in the toroidal source model) for the given best-fit values of the source magnetic field $B$.}
\label{MCMCfits}
\renewcommand{\arraystretch}{1.5}
\begin{tabular}{lccccccc}
\hline\hline
Object & SB & $L$, $R_*$ & $B$, G & $\theta$, degrees & $\log_{10}n_{\mathrm{b}}$, $\textrm{cm}^{-3}$ & $\delta$ & $b$\\
\hline
LP 349-25 & 3 & $12.78$ & $0 - 215$ (10.96) & $80$ & $1.94^{+1.45}_{-1.21}$ (4.08) & $1.71^{+0.48}_{-0.48}$ (1.94) & $1.00^{+1.01}_{-0.54}$\\
2M 0036 & 1 & $3.85$ & $173.37^{+43.84}_{-30.69}$ (174.31) & $141.21^{+4.04}_{-4.04}$ (141.45) & $4.31^{+0.86}_{-0.86}$ (4.50) & $2.43^{+0.34}_{-0.34}$ (2.49) & $5.17^{+2.99}_{-2.64}$\\
NLTT 33370 & 1 & $10.49$ & $18.07^{+9.60}_{-5.28}$ (18.26) & $151.59^{+3.18}_{-3.64}$ (151.52) & $7.35^{+0.86}_{-1.00}$ (7.69) & $2.85^{+0.28}_{-0.37}$ (2.97) & $2.31^{+2.01}_{-1.21}$\\
TVLM 513 & 1 & $4.60$ & $120.10^{+22.22}_{-18.18}$ (122.45) & $80$ & $4.37^{+0.65}_{-0.65}$ (4.45) & $2.52^{+0.30}_{-0.30}$ (2.56) & $7.79^{+7.86}_{-4.84}$\\
LSR 1835 & 2 & $2.55$ & $353.61^{+63.16}_{-39.47}$ (355.25) & $35.52^{+4.51}_{-3.22}$ (35.49) & $3.15^{+0.42}_{-0.46}$ (3.21) & $1.65^{+0.19}_{-0.21}$ (1.67) & $73.12^{+266.48}_{-60.17}$\\
\hline
\end{tabular}
\end{table*}

Radio-emitting ultracool dwarfs are expected to possess the magnetic fields with the strengths of a few thousand Gauss at the surface level, therefore, we require our simulations to be consistent with that estimation. Table \ref{MCMCfits} lists the results of the MCMC analysis for several datasets (one dataset for each object). The characteristic source sizes $L$ are assumed to be fixed during simulations; the source size for each object was chosen to ensure that the equivalent surface magnetic field strength $B_0$ (in the toroidal source model) for the best-fit set of parameters equals $B_0=3500$ G. For the other model parameters, Table \ref{MCMCfits} reports their most probable values together with the uncertainty limits (at 75\% confidence level). We can see that the considered ultracool dwarfs seem to possess very diverse magnetospheres, since the estimated emission source parameters (including the source size and magnetic field) are very different for different objects. For most of the objects, the MCMC analysis is able to constrain effectively the model parameters, the uncertainty ranges are relatively narrow, and the maxima of the posterior probability distributions coincide with the least-squares best-fit parameter combinations; the only exception is the dwarf LP 349-25 which will be discussed later.\

\begin{table*}
\caption{Parameters of the gyrosynchrotron emission source fitting the observations for each object and scheduling block (SB): the characteristic source size $L$, the magnetic field strength $B$ and viewing angle $\theta$, the concentration of energetic electrons $n_{\mathrm{b}}$ and their spectral index $\delta$; the equivalent major radius of the toroidal source model $R_{\mathrm{c}}$ is also presented. The parameter ranges correspond to the equivalent surface magnetic field strengths of $B_0=2000-5000$ G.}
\label{GSfits}
\renewcommand{\arraystretch}{1.4}
\begin{tabular}{lccccccc}
\hline\hline
Object & SB & $L$, $R_*$ & $B$, G & $\theta$, degrees & $\log_{10}n_{\mathrm{b}}$, $\textrm{cm}^{-3}$ & $\delta$ & $R_{\mathrm{c}}$, $R_*$\\
\hline
LP 349-25  & 1 & $9.42-13.41$  & $13.69-13.83$   & $80$            & $3.85-3.34$ & $1.75-1.73$ & $4.18-5.65$\\
LP 349-25  & 3 & $11.10-13.11$ & $9.05-14.10$    & $80$            & $4.41-3.85$ & $1.94-1.92$ & $4.80-5.54$\\
LP 349-25  & 4 & $9.82-12.85$  & $12.34-15.49$   & $80$            & $3.45-2.93$ & $1.56-1.54$ & $4.33-5.44$\\
2M 0036    & 1 & $3.30-4.25$   & $132.81-204.68$ & $147.41-137.73$ & $5.42-3.97$ & $2.66-2.38$ & $1.96-2.30$\\
2M 0036    & 2 & $3.00-3.99$   & $156.42-231.72$ & $156.08-147.99$ & $5.39-4.10$ & $2.49-2.31$ & $1.86-2.21$\\
NLTT 33370 & 1 & $8.74-11.38$  & $16.49-21.07$   & $157.72-148.04$ & $7.59-7.34$ & $2.72-2.97$ & $3.93-4.90$\\
TVLM 513   & 1 & $4.02-5.03$   & $91.82-144.52$  & $80$            & $5.13-4.04$ & $2.68-2.49$ & $2.22-2.58$\\
LSR 1835   & 1 & $2.06-2.81$   & $275.54-433.33$ & $36.85-51.22$   & $3.39-2.50$ & $1.41-1.34$ & $1.54-1.79$\\
LSR 1835   & 2 & $2.07-2.88$   & $275.25-418.18$ & $28.77-40.20$   & $3.87-2.79$ & $1.72-1.61$ & $1.54-1.81$\\
\hline
\end{tabular}
\end{table*}

Considering the equivalent magnetic field strength at the surface level $B_0$ as a free parameter, we now allow it to vary within the range of $B_0=2000-5000$ G; the corresponding ranges of the fitted model parameters (obtained using the least-squares fitting) are listed in Table \ref{GSfits}. The resulting best-fit emission spectra are shown in Figure \ref{SED} together with the observed data; for each object, we plot two model spectra corresponding to two different best-fit sets of parameters (for equivalent $B_0=2000$ and 5000 G, respectively). Note that the parameter ranges presented in Table \ref{GSfits} are not the uncertainty limits. Instead, they reflect the fact that the emission model parameters cannot be determined uniquely but only as functions of some free parameter; at the same time, they are correlated with each other. The actual uncertainty limits are proportional to those listed in Table \ref{MCMCfits}.\

\subsubsection{Modelling results: discussion}\label{mrdiscussion}\

Here we briefly discuss the observed properties of the quiescent emission and the inferred parameters of the emission sources for different ultracool dwarfs.\

{\sl LP 349-25}: The spectral peak of the emission seems to be located below 2 GHz, i.e., beyond the spectral range of our observations. The lack of information about the spectral peak reduces the efficiency of the gyrosynchrotron diagnostics. MCMC analysis yields poorly constrained results even for a fixed source size $L$, because the posterior probability distributions of the model parameters (especially of the magnetic field strength $B$) are too broad. Thus, the results of the least-squares fitting (presented in Tables \ref{MCMCfits} and \ref{GSfits}) are not so reliable as well. Nevertheless, considering them as typical examples of suitable parameter sets, we can conclude that the magnetic field in the source region should be relatively weak (with the strength of about 10 G), which corresponds to an extended emission source (with equivalent $R_{\mathrm{c}}\simeq 5R_*$) and a relatively low concentration of energetic electrons ($n_{\mathrm{b}}\simeq 10^3-10^4$ $\textrm{cm}^{-3}$). The electron spectral index can be determined more accurately and implies a rather hard spectrum ($\delta\simeq 1.75$); similar estimations of $\delta$ follow from the observations of \citet{osten09}. LP 349-25 is a binary system consisting of two nearly identical ($\sim$M8) brown dwarfs \citep{dupuy10}. Currently we cannot tell which component (or both of them) produces the radio emission \citep[see also][]{osten09}; the above estimations correspond to the case of a single emission source. Wide separation of the binary components (${\simeq 1.94}$ AU, i.e., much larger than the estimated size of the emission source) excludes interaction of their magnetospheres. \citet{dupuy10} estimated the inclination of the orbital plane of LP 349-25 as $i\simeq 117^{\circ}$; assuming spin-orbit alignment, we can take this value as an (approximate) inclination of the rotation axes of the system components relative to the line-of-sight. This inclination is rather high and hence non-detection of circular polarisation agrees with an assumption that the magnetic field of the radio-emitting dwarf (or dwarfs) is nearly axisymmetric; in this case, the total (i.e., integrated over both magnetic hemispheres) Stokes $V$ flux should be low. The observations at three different times provide consistent results.\

{\sl 2M 0036}: The spectrum shape is consistent with the gyrosynchrotron model, with the spectral peak at about $3-4$ GHz. In addition, circular polarisation was reliably detected in a broad frequency range, which allows us to estimate the magnetic field inclination. The inferred emission source size (with equivalent $R_{\mathrm{c}}\simeq 2R_*$) is comparable to that at Jupiter, but the magnetic field strength ($B\simeq 150-200$ G) and electron fluxes (with $n_{\mathrm{b}}\simeq 10^4-10^5$ $\textrm{cm}^{-3}$) are much higher than in the Jovian radiation belts. High degree of circular polarisation requires an oblique magnetic field with the viewing angle of $\theta\simeq 140^{\circ}-150^{\circ}$ (or $30^{\circ}-40^{\circ}$ if we do not consider the polarisation sign). On the other hand, the observed rotation velocity of $v\sin i\simeq 35.12\pm 0.57$ km $\textrm{s}^{-1}$ \citep{blake10}, rotation period of $T\simeq 3.08\pm 0.05$ h \citep{hallinan08} and radius of $0.88\pm 0.05$ $R_{\mathrm{J}}$ \citep{sorahana13} imply that the rotation axis of this dwarf is nearly perpendicular to the line of sight ($i\simeq 65^{\circ}-90^{\circ}$). Therefore, we conclude that the magnetic field of 2M 0036 should be considerably asymmetric with respect to the rotation axis (e.g., a highly tilted dipole). The observations at two different times provide consistent results. One may expect that rotation of a tilted magnetic dipole would result in oscillating sign of the Stokes $V$ signal, and the observations of \citet{hallinan08} indeed show evidence of such behaviour. However, we do not analyse the detailed temporal evolution of the quiescent emission here (due to insufficient observation coverage). In addition, two observational sessions on 2014-10-07 and 2014-10-08 corresponded to similar rotation phases of 2M0036, therefore, the observed spectra (including the polarisation sign) were similar as well.\

{\sl NLTT 33370}: The spectral peak of the emission is located at low frequencies (about 2 GHz) which indicates a weak magnetic field. On the other hand, the emission intensity is extremely high: up to $2\times 10^{10}$ Jy when normalized to 1 AU distance, which corresponds to the spectral luminosity (assuming isotropic emission) of up to $5\times 10^{14}$ erg $\textrm{s}^{-1}$ $\textrm{Hz}^{-1}$. Such combination can be achieved only due to the large source size and/or high concentration of emitting particles. Fitting infers the magnetic field strength of about 20 G, the emission source size equivalent to $R_{\mathrm{c}}\simeq 4.5R_*$, and the concentration of energetic electrons exceeding $10^7$ $\textrm{cm}^{-3}$. The circular polarisation degree demonstrates a strong frequency dependence: it is relatively high (up to 40\%) near the spectral peak, but rapidly decreases at higher frequencies; this behaviour cannot be reconciled with the gyrosynchrotron model (cf. the model spectra in Figure \ref{SED}). We suggest that the observed Stokes $V$ spectrum can be formed due to the source inhomogeneity in a dipolar magnetosphere: near the spectral peak (where the optical depth is close to unity) the observed emission is produced in a relatively narrow layer with a constant sign of the magnetic field. On the other hand, in the optically thin range we observe the emission integrated over the entire magnetosphere (i.e., including contributions of opposite magnetic hemispheres) which reduces the total polarisation. Our simplified homogeneous source model cannot account for these effects, therefore, as said above, in the fitting procedure the polarisation non-detections have been treated as missing data rather than as intrinsically zero polarisation. To provide the observed polarisation degree (at low frequencies), the magnetic field should be nearly parallel to the line-of-sight: $\theta\simeq 150^{\circ}-160^{\circ}$, or $\theta=20^{\circ}-30^{\circ}$ if we do not consider the polarisation sign. 

NLTT 33370 is a binary consisting of two $\sim$M7 ultracool dwarfs \citep{mclean11, schlieder14, williams15,dupuy16}. Recently, \citet{forbrich16} have resolved the binary with VLBA observations and found only the secondary component (NLTT 33370 B) to be active with the average flux density of 666 $\mu$Jy at 7.2 GHz, and gave an upper limit of 22 $\mu$Jy for NLTT 33370 A. Therefore, our above estimations refer to that active component (NLTT 33370 B). The rotation axis of the active component of NLTT 33370 seems to be strongly inclined relative to the line of sight; \citet{mclean11} estimated the inclination as $i\gtrsim 70^{\circ}$ (although they attributed the radio emission to the primary component). Therefore, like in the case of 2M 0036, the magnetic field of the active component of NLTT 33370 should be strongly tilted; a similar conclusion was made by \citet{mclean11}. Wide separation of the binary components ($\sim$2 AU, i.e., much larger than the estimated size of the emission source) excludes interaction of their magnetospheres. Note that our results for NLTT 33370 differ somewhat from the results presented by \citet{mclean11} and \citet{williams15}: in our observations, the Stokes $I$ spectrum at ${\gtrsim 4}$ GHz decreases with frequency much faster than in the cited works. In addition, \citet{mclean11} and \citet{williams15} detected noticeable Stokes $V$ signals at the frequencies above 4 GHz and even up to 8.5 GHz. These differences cannot be fully explained by the lower temporal resolution and shorter duration of our observations, therefore we attribute them to long-term variations of the emission source (e.g., changing parameters of the energetic electrons).\

\begin{figure}
\includegraphics{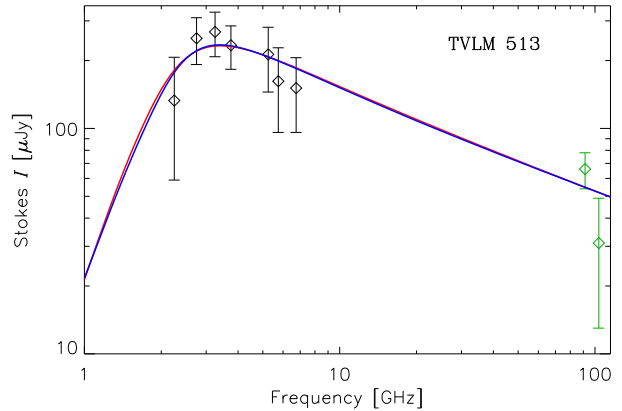}
\caption{Combined non-simultaneous broadband spectrum of TVLM 513. Black points: VLA observations (2015-02-06, this work); green points: ALMA observations \protect\citep[2015-04-03,][]{williams-ALMA}. The continuous curves represent the gyrosynchrotron fits (same as in Figure \protect\ref{SED}).}
\label{VLA_ALMA}
\end{figure}

{\sl TVLM 513}: The spectrum shape is consistent with the gyrosynchrotron model, with a well-defined peak at about $3.0-3.5$ GHz. Fitting infers the emission source parameters similar to those for 2M 0036 (except of the viewing angle): $B\simeq 100$ G, $R_{\mathrm{c}}\simeq 2.4R_*$, $n_{\mathrm{b}}\simeq 10^4-10^5$ $\textrm{cm}^{-3}$, $\delta\simeq 2.6$. The inferred magnetic field strength falls into the ranges found by \citet{osten06b} and \citet{berger08}. The rotation axis of TVLM 513 seems to be nearly perpendicular to the line-of-sight \citep[$i\simeq 70^{\circ}-85^{\circ}$, according to estimations of][]{hallinan06, berger08, milespaez15}, therefore, like in the case of LP 349-25, non-detection of circular polarisation agrees with the assumption that the magnetic field is nearly axisymmetric. It is interesting to note that recently \citet{williams-ALMA} detected emission from TVLM 513 in the millimetre range with ALMA. Figure \ref{VLA_ALMA} presents the combined spectrum in the $2-100$ GHz range (although we highlight that the VLA and ALMA observations were not simultaneous). To reproduce this spectrum with the gyrosynchrotron model, we would need a harder electron spectrum ($\delta\simeq 1.75$) and a lower electron density ($n_{\mathrm{b}}\simeq 10^3$ $\textrm{cm}^{-3}$) than for the VLA observations alone. The required source size and magnetic field strength ($R_{\mathrm{c}}\simeq 2.3R_*$ and $B\simeq 140$ G) are close to the above-mentioned values. Thus, it is likely that a few data points below 10 GHz are actually insufficient to constrain properly the electron spectral index. As shown by \citet{kuznetsov11}, in an inhomogeneous emission source the spectral index of the optically thin gyrosynchrotron emission is frequency-dependent and approaches a constant only at the frequencies far above the spectral peak; hence estimating the electron spectral index using the emission spectral index in a limited frequency range may be unreliable. An alternative explanation is that the discrepancies between the VLA and ALMA observations are caused by long-term variability (at the timescales of ${\sim 2}$ months).\

{\sl LSR 1835}: The spectra, obtained at two different epochs look similar and deviate noticeably from the predictions of the gyrosynchrotron model. A possible interpretation is that the intense (but weakly polarised) emission at $2.5-3.0$ GHz is produced due to a coherent mechanism; e.g., it may be a maser emission scattered and depolarised during propagation \citep[as suggested by][]{hallinan08}. If we omit the data point at $2.5-3.0$ GHz, the fitting procedure infers a relatively compact emission source with strong magnetic field (see Figure \ref{SED} and Tables \ref{MCMCfits}-\ref{GSfits}). Like in the case of NLTT 33370, we do not consider the polarisation non-detections in the fitting procedure, although the same interpretation of them (by the source inhomogeneity) is questionable since the polarised signal is below the detectability threshold even near the suggested spectral peak. At the same time, we think that the data in the $3-7$ GHz range are insufficient to constrain the gyrosynchrotron source parameters for these particular spectra (both due to the small number of data points and due to the relatively narrow spectral range). In addition, the maser emission may make a contribution at higher frequencies as well (as suggested, e.g., by detection of flaring emission from this dwarf). Therefore, we conclude that the obtained gyrosynchrotron fits (including the estimations of the viewing angle) for the observed emission from LSR 1835 are not reliable enough; we do not discuss them in detail here.\

\begin{figure}
\centering
\includegraphics{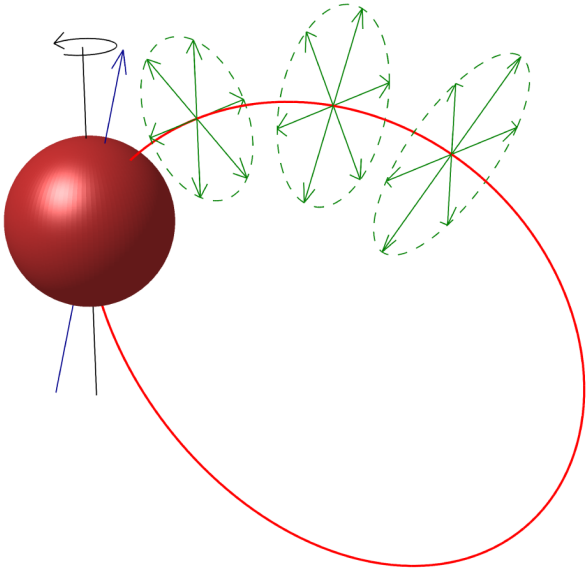}
\caption{A schematic model of the emission source structure for the dwarf LSR 1835. The blue arrow represents the magnetic dipole direction, the red line is a selected ``active'' magnetic field line, and green arrows represent the directions of the radio emission. The Figure proportions are chosen for illustrative purposes only, i.e., they do not correpond to the model parameters used in the simulations.}
\label{DwarfModel}
\end{figure}

\subsection{Flaring radio emission from LSR 1835}\label{Fsim}
\subsubsection{Source model}\

We simulate the flaring radio emission from LSR 1835 using a model similar to that in the work of \citet{kuznetsov12}. Namely, the magnetic field of the dwarf is modelled by a tilted dipole; the emission is assumed to be produced at a few selected ``active'' magnetic field lines (see Fig. \ref{DwarfModel}). We assume that the emission is produced due to the shell-driven electron-cyclotron maser instability \citep[][and references therein]{treumann06, hess11, kuznetsov12}, i.e., its frequency nearly coincides with the electron-cyclotron frequency at the source, and its direction is perpendicular to the magnetic field vector at the source (a possible refraction during propagation is neglected). The ``active'' magnetic field lines are fixed in the rotating frame of the dwarf, therefore, as the dwarf rotates, we can observe the radio emission when the radio beam (at a given frequency, which corresponds to a certain height of the emission source above the dwarf surface) is directed towards the Earth.\

As was shown by \citet{kuznetsov12}, the above model itself is able to reproduce the narrowband periodic radio bursts with fast frequency drift; the dynamic spectrum containing several narrowband bursts (Fig. \ref{lsr-flare}) implies that the model should contain the same number of ``active'' field lines. However, to explain the faint slowly-drifting broadband feature (whose frequency drift direction is opposite to that of the narrowband bursts), we need to use a more elaborate model, i.e., to consider the limited (and variable) height extent of the emission source. Therefore, we describe the emission intensity from a given ($j$-th) magnetic field line by the expression

\begin{equation}
I_j=a_j(\lambda, r)\exp\left[-\frac{(\theta-\pi/2)^2}{\Delta\theta_j^2}\right],
\end{equation}

where $\theta$ is the angle between the emission direction and the local magnetic field and $\Delta\theta_j$ is the radio beam half-width; this expression means that the emission is produced in a narrow range of angles around the direction perpendicular to the local magnetic field. The factor $a_j(\lambda, r)$ describes the dependence of the emission intensity on the coordinates of the emission source: the magnetic longitude $\lambda$ (which is calculated relative to the plane containing the rotation axis and the dipole axis and increases in the direction of the dwarf rotation) and the distance from the dipole centre $r$. We have chosen the following expression to describe this dependence:

\begin{equation}
a_j(\lambda, r)=A_j\exp\left\{-\frac{[r-r_0(\lambda)]^2}{\Delta r^2(\lambda)}\right\},
\end{equation}

where $A_j$ is the maximum emission intensity for the given magnetic line, and $r_0(\lambda)$ and $\Delta r(\lambda)$ are the typical height and height extent of the emission source, respectively (both may be dependent on the magnetic longitude). Other parameters describing the emission source model are the magnetic field strength at the magnetic pole $B_0$ (assuming that the dipole centre coincides with the dwarf centre), the dipole inclination relative to the rotation axis $\delta$, the radii (or the $L$-shell numbers) of the ``active'' field lines $L_j$, and the inclination of the dwarf rotation axis relative to the line-of-sight $i$.

\subsubsection{Simulation results}\

Since even the above-described (oversimplified) model has a lot of free parameters, it is not currently possible to find a unique set of the model parameters that fit the observations; the low signal-to-noise ratio and limited duration of the observations (which does not allow us to study the possible periodicity of the emission) hamper the quantitative analysis as well. Therefore, our aim was to find one feasible model that would agree with the observations. Namely, such model should reproduce the main features visible in the dynamic spectrum in Fig. \ref{lsr-flare}: the faint broadband slow-drifting burst and the bright narrowband fast-drifting bursts, their frequency drifts and frequency extents.

\begin{figure}
\begin{center}
\includegraphics{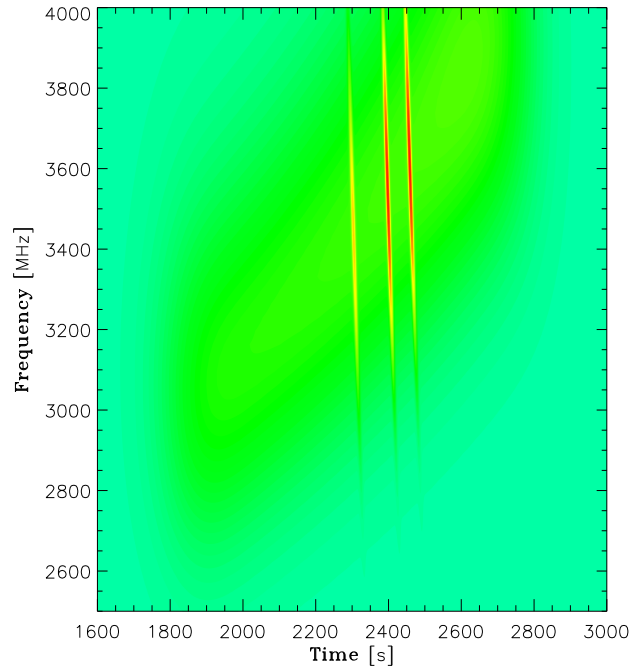}
\end{center}
\caption{Simulated dynamic spectrum of the flaring radio emission (Stokes $V$) from LSR 1835.}
\label{ModelSpectrum}
\end{figure}

The simulation results are shown in Fig. \ref{ModelSpectrum}. In this model, we use the estimation of $B_0=3625$ G for the magnetic field strength at the magnetic pole \citep[as follows from the observations of][]{berdyugina15}. All ``active'' magnetic field lines are assumed to have the same $L$-shell number of $L=80$, the inclination of the dwarf rotation axis relative to the line-of-sight is $i=77^{\circ}$, and the dipole tilt relative to the rotation axis is $\delta=10^{\circ}$. We have found that the above parameters provide an acceptable fit to the observed frequency drift rates of the fast-drifting bursts (i.e., they reproduce both the drift rate value and the fact that the drift rate seems to be constant in a broad range of frequencies). The above parameters imply that the magnetic field is highly axisymmetric and the radio emission is produced at high magnetic latitudes (like, e.g, at the Earth). 

Three bright fast-drifting narrowband bursts are modelled by using three ``active'' field lines at $\lambda_j=0^{\circ}$, $4^{\circ}$, and $10^{\circ}$ with a very narrow emission directivity of $\Delta\theta_j=0.05^{\circ}$. The faint broadband burst is modelled by 15 ``active'' field lines evenly distributed in the range of longitudes from $\lambda=-20^{\circ}$ to $\lambda=36^{\circ}$ (i.e., with the total longitude extent of the ``active'' sector of about $55^{\circ}$); the emission sources at these lines are assumed to have a relative broad directivity of $\Delta\theta_j=1^{\circ}$ and the amplitude $A_j$ five times lower than at the field lines corresponding to the fast-drifting bursts. We assume that the typical height of the emission region $r_0$ (relative to the dipole centre) varies linearly with the magnetic longitude $\lambda$ from $r_0=1.36R_*$ at $\lambda=-20^{\circ}$ to $r_0=1.50R_*$ at $\lambda=36^{\circ}$, where $R_*\simeq 70\,000$ km is the dwarf radius; the parameter $\Delta r$ is taken to be $0.05R_*$, i.e., the height extent of the emission region is of about $2\Delta r\simeq 0.1R_*\simeq 7000$ km. The mentioned dependence of the height and height extent of the radio emission source on the magnetic longitude is applied to all ``active'' magnetic field lines, which allows us to reproduce both the frequency drift of the faint broadband feature and the frequency extent of the broadband feature and the narrowband fast-drifting bursts. The described model contains a number of numerical parameters; we remind, however, that they are largely illustrative and represent only one possible combination fitting the observations. Investigating the confidence limits of the model parameters is beyond the scope of this work.

\subsubsection{Comparison with other simulations}\

As has been said above, the model of the flaring emission used in this work is essentially the same as in the paper of \citet{kuznetsov12}, i.e., it is based on the assumption of a global dipole-like magnetic field and a number of ``active'' longitudes. The difference is that the model of \citet{kuznetsov12} was able to reproduce only the ``skeletons'' of the bursts in the dynamic spectra, while now we consider also a possible dependence of the emission intensity on the source height and hence on the frequency. On the other hand, we consider here only the shell-driven maser emission that is assumed to be strictly perpendicular to the source magnetic field, while \citet{kuznetsov12} analysed also the loss-cone-driven emission produced in oblique directions. 

It is interesting to note that \citet{lynch15} concluded that the model of \citet{kuznetsov12} was unable to reproduce their observations (in application to the dwarfs 2M J0746+2000 and TVLM 513). Instead, they proposed a model with multiple subsurface magnetic dipoles, i.e., with an essentially multipolar magnetic field; each local dipole was associated with a single ``active'' magnetic loop. In contrast, we have found here that the model with a purely dipolar field is able to reproduce successfully the dynamic spectrum of LSR 1835, therefore, more complicated models are not needed (at least, in this case).

\section{Conclusions}\

We have carried out observations of six UCDs in the S and C bands using the VLA and found that five out of six sources were detected at a significant level. We have analysed the origins of the quiescent radio emission via the broadband spectra. Modelling was done with both MCMC analysis and least square fitting, showing very similar results. The spectrum shape and degree of circular polarisation of the quiescent emission from four of the five observed and detected sources are found to be consistent with the predictions for the gyrosynchrotron mechanism. Based on the model we have given a set of parameters for the magnetic field and energetic electrons for each dwarf.\

A flare-like feature which was 100\% circularly polarised emission was detected from LSR J1835+3259. The event has a broad component with a frequency drift of approximately 1 MHz s$^{-1}$, and narrow components which show a drift of approximately -30 MHz s$^{-1}$. Bursts with high brightness temperature and polarisation degree from UCDs have been found to exhibit emission properties that are similar to the auroral radio emissions of the magnetised planets of the Solar system \citep{hallinan15}. Although this is not the first time that such events have been observed from UCDs, it is one of the few bursts which have shown such clear frequency drifts. As far as we know this is the first detection of flaring (100\% circularly polarised) emission from a UCD at frequencies below 4GHz. Also, this is the first time such flaring event on a UCD demonstrates both positive and negative frequency drifts.\

Using simulations, we have come up with a possible model that fits the observed characteristics of the flaring emission,  including the frequency drifts. We found that if we fix several model parameters (based on physical characteristics of similar stars), we can match our observations with emission coming from a narrow sector of active longitudes and the dwarf's magnetic field of a tilted dipole. The model may differ from the observations not due to the magnetic field structure, but the height distribution of energetic electrons in the source region. The variable height of the radio-emitting region can be explained by different reasons, for example, if the magnetic dipole is not only tilted relative to the dwarf rotation axis, but also offset relative to the dwarf centre (like, e.g., at Neptune). Other explanations include an influence of the higher-order (non-dipole) magnetic field components, centrifugal force effects, etc. However, extracting parameters from this model is more difficult due to the underlying degeneracy.\

Radio observations of UCDs are entering a new era with the improvement of existing radio arrays such as the VLA which allow the monitoring of single sources over a much wider bandwith and temporal resolution than was previously possible. This is especially important for studies of UCDs where the radio burst/pulse can be of short duration and can change frequency over a short timescale. In addition, with new facilities in development, such as SKA, the observations will be able to cover a wider spectral range and will give higher sensitivity, making it possible to detect even fainter objects over a wide frequency range.\

\section{Acknowledgements}\

Research at Armagh Observatory is grant-aided by the Northern Ireland Department for Communities and via the Science and Technology Facilities Council consolidated grant. A.A. and Y.M. acknowledge the support of the University of Sofia - Science Foundation grant No 81/2016. We thank Gregg Hallinan, Stephen Bourke and Colm Coughlan for help with the data reduction. We also thank the referee for a constructive and useful report. This research has made use of the SIMBAD database, operated at CDS, Strasbourg, France. The VLA is operated by the National Radio Astronomy Observatory, a facility of the National Science Foundation operated under cooperative agreement by Associated Universities, Inc. \

\bsp

\bibliographystyle{mn2e}

\bibliography{vla_v5}

\label{lastpage}

\end{document}